\documentclass[twoside,11pt]{article}
\usepackage{jair, theapa, rawfonts}

\usepackage{amsfonts,amssymb,amsmath}
\usepackage{verbatim}
\usepackage{color}
\usepackage{float}
\usepackage{times}
\usepackage{graphicx}
\usepackage{latexsym}
\usepackage{helvet}
\usepackage{breqn}

\usepackage{epic}
\usepackage[small]{caption}

\newtheorem{definition}{Definition}
\newtheorem{proposition}{Proposition}
\newtheorem{theorem}{Theorem}
\newtheorem{lemma}{Lemma}

\newcommand{\bull}{\rule{.85ex}{1ex} \par \bigskip}

\newenvironment{altproof}{\noindent {\bf Proof:\ }}{\bull}
\newcommand{\tuple}[1]{\ensuremath{\langle #1 \rangle}}

\thicklines 
\newsavebox{\varthree}
\savebox{\varthree}(20,40){
\begin{picture}(20,40)(0,0)
\put(10,20){\oval(18,38)} 
\put(10,10){\makebox(0,0){{\large $\bullet$}}}
\put(10,20){\makebox(0,0){{\large $\bullet$}}} 
\put(10,30){\makebox(0,0){{\large $\bullet$}}}
\end{picture}
}
\newsavebox{\vartwo}
\savebox{\vartwo}(20,40){
\begin{picture}(20,40)(0,0)
\put(10,20){\oval(18,38)} \put(10,10){\makebox(0,0){{\large $\bullet$}}}
\put(10,30){\makebox(0,0){{\large $\bullet$}}}
\end{picture}
}
\newsavebox{\varone}
\savebox{\varone}(20,40){
\begin{picture}(20,40)(0,0)
\put(10,20){\oval(18,28)} \put(10,20){\makebox(0,0){{\large $\bullet$}}}
\end{picture}
}
\thicklines 
\newsavebox{\varthreebigbis}
\savebox{\varthreebigbis}(20,45){
\begin{picture}(20,45)(0,0)
\put(15,20){\oval(18,38)} 
\put(10,10){\makebox(0,0){{\Large $\bullet$}}}
\put(10,20){\makebox(0,0){{\Large $\bullet$}}} 
\put(10,30){\makebox(0,0){{\Large $\bullet$}}}
\end{picture}
}
\newsavebox{\varthreebig}
\savebox{\varthreebig}(20,40){
\begin{picture}(20,40)(0,0)
\put(10,20){\oval(18,38)} 
\put(10,10){\makebox(0,0){{\Large $\bullet$}}}
\put(10,20){\makebox(0,0){{\Large $\bullet$}}} 
\put(10,30){\makebox(0,0){{\Large $\bullet$}}}
\end{picture}
}
\newsavebox{\vartwobig}
\savebox{\vartwobig}(20,40){
\begin{picture}(20,40)(0,0)
\put(10,20){\oval(18,38)} 
\put(10,10){\makebox(0,0){{\Large $\bullet$}}}
\put(10,30){\makebox(0,0){{\Large $\bullet$}}}
\end{picture}
}
\newsavebox{\vartwobigbis}
\savebox{\vartwobigbis}(20,40){
\begin{picture}(20,40)(0,0)
\put(5,20){\oval(18,38)} 
\put(10,10){\makebox(0,0){{\Large $\bullet$}}}
\put(10,30){\makebox(0,0){{\Large $\bullet$}}}
\end{picture}
}
\newsavebox{\varonebig}
\savebox{\varonebig}(20,40){
\begin{picture}(20,40)(0,0)
\put(10,20){\oval(18,28)} \put(10,20){\makebox(0,0){{\Large $\bullet$}}}
\end{picture}
}
\thinlines 

\begin{document}

\title{Variable elimination in binary CSPs}

\author{\name Martin C. Cooper \email cooper@irit.fr \\
       \addr IRIT, University of Toulouse III, \\
       31062 Toulouse, France 
       \AND
       \name Achref {El~Mouelhi} \email elmouelhi.achref@gmail.com \\
       \addr H \& H: Research and Training, \\
       13015 Marseille, France
       \AND
       \name Cyril Terrioux \email cyril.terrioux@lis-lab.fr \\
       \addr Aix Marseille Univ, Universit\'e de Toulon, \\
       CNRS, LIS, Marseille, France}


\maketitle

\begin{abstract}
We investigate rules which allow variable elimination in binary CSP (constraint
satisfaction problem) instances while conserving  
satisfiability. We study variable-elimination rules based on the
language of forbidden patterns enriched with counting and quantification over variables and values.
We propose new rules and compare them, both theoretically and experimentally. 
We give optimised algorithms to apply these rules and show that each define a novel
tractable class. Using our variable-elimination rules in preprocessing allowed us
to solve more benchmark problems than without.
\end{abstract}

\section{Introduction}

Constraint satisfaction provides a generic model 
for many NP-hard problems encountered in fields such as artificial intelligence, bioinformatics and operations research.
In this paper, we study binary CSP instances, 
in which each constraint concerns at most
two variables. It is well known that all CSP instances can be expressed as binary instances,
via the dual encoding~\cite{dechter89:tree} or the hidden variable encoding~\cite{rossi90:equivalence}. 

Since the binary CSP is NP-complete, it is of practical interest to find polynomial-time operations
which reduce the size of the search space. 
One obvious way to reduce search space size is by
variable elimination. 

Variable elimination is classic in those families of constraint problems in which variables can be eliminated
without changing the nature of the constraints: we can cite Gaussian elimination in systems of linear
equations over a field~\cite{Schrijver} 
or variable-elimination resolution in boolean formulae in CNF~\cite{DBLP:conf/sat/SubbarayanP04}. 
Indeed, any variable $x_i$ can be eliminated
from a \emph{general-arity} CSP instance by joining all constraints whose scope includes $x_i$ and projecting
the resulting relation $R$ with scope $Y$ onto the variables 
$Y \setminus \{x_i\}$~\cite{DBLP:journals/ai/Dechter99,DBLP:journals/constraints/LarrosaD03}. Call this relation $R^{-x_i}$. 
Unfortunately, this often introduces a high-arity constraint and this can be counterproductive in terms
of both memory and time. Under certain conditions, a binary CSP instance will remain binary after this 
join-and-project variable elimination of $x_i$. For example, this is clearly the case if $x_i$ is constrained by
only two other variables since, in this case, $R^{-x_i}$ is binary. 
A more interesting case is when all constraints with $x_i$ in their scope
share a majority polymorphism since, in this case, the relation $R^{-x_i}$ 
is equivalent to the join of its binary projections~\cite{DBLP:journals/ai/JeavonsCC98}. 
Fourier's algorithm for variable elimination applied to 
a system of binary linear inequalities~\cite{DBLP:reference/fai/Koubarakis06,Schrijver}
can be viewed as just one example of this general rule, since binary linear inequalities
are all closed under the majority polymorphism \emph{median}. Another interesting case is when there is a functional
constraint of the form $x_i = f(x_j)$ (where $f$ is a function) for some other variable $x_j$: 
the relation $R^{-x_i}$ is then equivalent
to the join of its projections onto the pairs of variables $(x_j,x_k)$ ($k \neq i,j$)~\cite{DBLP:journals/tplp/ZhangY11}.

Unfortunately, the introduction of a large number of new constraints, even if they are still  binary, may again
be counterproductive. Therefore, we concentrate in this paper on rules which do not introduce new constraints when 
a variable is eliminated.

Various rules have been found which allow the elimination
of a variable 
without introducing new constraints and without changing the satisfiability of the instance~\cite{ve,beyond,Cooper10:btp}. 
Such rules were used, for example, in the deep optimisation solution to the spectrum repacking problem~\cite{deepOpt}.
Discovery of new variable-elimination rules may have not only practical but also theoretical applications.
For example, simple rules for variable or value elimination are used by Beigel and Eppstein~\cite{DBLP:conf/focs/BeigelE95}
in their algorithms with low worst-case time bounds for such NP-complete problems as 3-COLOURING and 3SAT:
these simplification operations are an essential first step before the use of decompositions into 
subproblems with smaller domains.
In the theory of fixed-parameter tractability, variable elimination is often an essential ingredient of
polynomial kernalisation algorithms. For example, in the Point Line Cover problem (find $k$ straight 
lines which cover $n$ points), if at least $k+1$ points lie on a line, then they can be effectively
eliminated since they must be covered by this line~\cite{DBLP:journals/talg/KratschPR16}. 
A form of variable elimination may also occur during the modelling phase. For example, in the modelling as a CSP
of the determination of the structure of a molecule from its chemical formula and other 
information obtained from nuclear magnetic resonance spectroscopy, 
the position of hydrogen atoms are not modelled since their positions are
uniquely determined by the multigraph of connections between the other atoms~\cite{DBLP:conf/setn/OmraniN16}.

We now define the notions that we will need in the rest of the paper.
\begin{definition}  \label{def:csp}
A \emph{binary CSP instance} $I=\tuple{X,\mathcal{D},R}$ comprises 
\begin{itemize}
\item  [$\bullet$] a set $X$ of $n$ variables $x_1,\ldots,x_n$,
\item  [$\bullet$] a domain $\mathcal{D}(x_i)$ for each variable $x_i$ ($i=1,\ldots,n$), and
\item  [$\bullet$] a binary constraint relation $R_{ij}$ for each pair of distinct variables $x_i, \! x_j$
($i, \! j \! \in \! \{1,\ldots,n\}$).
\end{itemize}
\end{definition}
For notational convenience, we assume that there is exactly one binary relation $R_{ij}$ for each pair of variables.
Thus, in the absence of an explicit constraint between $x_i$ and $x_j$, 
we define $R_{ij}$ to be $\mathcal{D}(x_i) \times \mathcal{D}(x_j)$.
Furthermore, $R_{ji}$ (viewed as a boolean matrix) is always the transpose of $R_{ij}$.
We say that $x_i$ \emph{constrains} $x_j$ if $R_{ij}$ is different from $\mathcal{D}(x_i) \times \mathcal{D}(x_j)$,
and we use $e$ to denote the number of pairs of variables $\{x_i,x_j\}$ such that $x_i$ constrains $x_j$.
An assignment $\tuple{v_1,\ldots,v_m}$ to variables $\tuple{x_{i_1},\ldots,x_{i_m}}$ is 
\emph{consistent} if $v_j \in \mathcal{D}(x_{i_j})$ (for $j=1,\ldots,m$) and 
$(v_j,v_k) \in R_{i_j,i_k}$ (for all $j,k$ such that $1 \leq j < k \leq m$).
A \emph{solution} to $I$ is a consistent assignment to all variables in $X$. For notational
convenience we can also view a solution as a mapping $s$ from $X$ to the union of the variable domains such
that $\tuple{s(x_1),\ldots,s(x_n)}$ is consistent.

It will sometimes be convenient to associate a binary CSP instance with its microstructure, a labelled graph
whose vertices are the variable-value assignments and which has positive and negative edges.
If $(v_i,v_j) \in R_{ij}$, we say that the assignments $\langle x_i,v_i \rangle$, $\langle x_j,v_j \rangle$ (or more
simply $v_i,v_j$) are \emph{compatible} and that
$v_iv_j$ is a \emph{positive edge}, otherwise $v_i,v_j$ are
\emph{incompatible} and $v_iv_j$ is a \emph{negative edge}. 
For simplicity of notation we can assume that variable domains are disjoint,
so that using $v_i$ as a shorthand for $\langle x_i,v_i \rangle$ is unambiguous.
We say that $v_i \in \mathcal{D}(x_i)$ has a \emph{support} at variable $x_j$ if 
there exists 
$v_j \in \mathcal{D}(x_j)$
such that $v_iv_j$ is a positive edge. 
A binary CSP instance $I$ is \emph{arc consistent} if for all pairs of distinct variables $x_i,x_j$, 
each $v_i \in \mathcal{D}(x_i)$ has a support at $x_j$.  
Arc consistency is ubiquitous in constraint solvers: it is 
applied both before and during search in binary CSPs since it can be established in $O(ed^2)$ time, 
where $e$ is the number of binary constraints and $d$ the maximum domain size~\cite{ac}.

In Section~\ref{sec:VErules} we introduce formally the notion of variable-elimination rule in binary CSPs
and give a known example (the $\exists$snake property) which we will compare theoretically and experimentally with novel 
variable elimination rules defined in this paper. Then, in Section~\ref{sec:DE-snake} we define a stronger rule, 
called DE-snake, which subsumes the $\exists$snake rule.
In Section~\ref{sec:triangle} we give the definition of
a simple variable elimination rule based on a triangle of variable-value assignments. It is well known
that the broken-triangle property~\cite{Cooper10:btp} is a variable-elimination rule. In Section~\ref{sec:from}
we generalise broken triangles to broken polyhedra. In Section~\ref{sec:existential} we give a family
of variable-elimination rules based on broken polyhedra of dimension $k$. On a more practical level,
in Section~\ref{faster_ve} we define a variable-elimination rule, based on the absence of broken
tetrahedra, which can be applied with the same worst-case time complexity as the broken-triangle rule
but is strictly stronger. In Section~\ref{sec:theory} we show that most of the rules we have
presented in this paper are theoretically incomparable. In Section~\ref{sec:experiments} we present the results
of our experimental trials on 3,557 benchmark instances. The variable-elimination rules
allowed us to solve more instances when they were applied in a preprocessing step, but we
recommend more research to better target those variables that are likely to be eliminated
before integrating these rules in a general-purpose solver. In Section~\ref{sec:tractability}
we show that each of the variable-elimination rules presented in this paper allows us to define a 
tractable class which can be recognised in polynomial-time.

\section{Variable-elimination rules}  \label{sec:VErules}

We study conditions under which a variable $x_i$ can
be eliminated from a binary CSP instance while conserving
satisfiability. A simple example of such a condition is that there exists a value
$v_i \in \mathcal{D}(x_i)$ which is compatible with all assignments to all
other variables. Clearly any solution $s$ to the instance $I'$ obtained
by eliminating $x_i$ can be extended to a solution to the original instance $I$ by setting 
$s(x_i)=v_i$. Another simple example is that the variable $x_i$ has
a singleton domain $\{v_i\}$. 
This second example demonstrates that when eliminating the variable
$x_i$ we need to retain the projections onto $X \setminus \{x_i\}$ of all
constraints whose scope includes $x_i$, since in this example we must
first eliminate from all domains $\mathcal{D}(x_j)$ ($j \neq i$) those
values that are not compatible with $\tuple{x_i,v_i}$. Thus, the instance
$I'$ obtained by \emph{eliminating a variable} $x_i$ from a binary CSP
instance $I$ is identical to $I$ except that (1) $\forall j \neq i$,
we have deleted from $\mathcal{D}(x_j)$ all values $v_j$ such that
$\tuple{x_j,v_j}$ has no support at $x_i$ in $I$, and (2) we have
deleted the variable $x_i$ and all constraints with $x_i$ in their scope.

We require the following formal definition in order to study provably-correct
variable-elimina\-tion rules~\cite{ve}.

\begin{definition} \label{def:var-elim}
A \emph{satisfiability-conserving variable-elimination condition} (or
a \emph{var-elim condition}) is a polytime-computable property $P(x_i)$
of a variable $x_i$ in a binary CSP instance $I$ such that when $P(x_i)$
holds the instance $I'$ obtained from $I$ by eliminating $x_i$
is satisfiable if and only if $I$ is satisfiable. Such a property
$P(x_i)$ is a \emph{solution-conserving variable-elimination condition}
(\emph{sol-var-elim condition}) if it is possible to construct a
solution to $I$ from any solution $s'$ to $I'$ in polynomial time.
\end{definition}

A sol-var-elim condition not only allows us to eliminate variables
while conserving satisfiability but also allows the polynomial-time
recovery of at least one solution to the original instance $I$ from a
solution to the reduced instance $I'$. All the var-elim properties
given in this paper are also sol-var-elim properties.

We end this section by giving an example of a known variable-elimination rule called the $\exists$snake rule.
It is based on forbidding a pattern of positive and negative edges (shown in Figure~\ref{fig:VEsnake}) on 
one value $v_i$ for $x_i$, the variable to be eliminated. In figures, broken lines represent negative edges
(incompatible pairs) and solid lines represent positive edges (compatible pairs).
A pattern is a (generally small) binary CSP instance in which the compatibility of certain values
(such as $v_i,v_k$ in Figure~\ref{fig:VEsnake}) may be left unspecified. 
A pattern $P$ occurs in a CSP instance $I$ if there is a homomorphism from $P$ to $I$ respecting variables and 
mapping positive edges to positive edges and negative edges to negative edges~\cite{ve}.
The $\exists$snake rule is one of the four variable-elimination rules based on forbidding an irreducible pattern
on the variable to be eliminated~\cite{ve}. Out of these four rules, we chose $\exists$snake to compare with the new
rules presented in this paper since, among these four rules, it would appear to be the most
promising in terms of time complexity and eliminating power.

\setlength{\unitlength}{1pt}
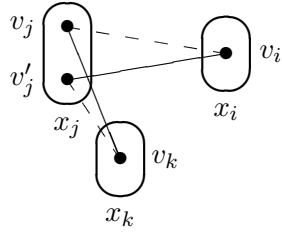
\begin{figure}
\centering

\begin{picture}(110,110)(-10,0)
\put(0,50){\usebox{\vartwo}} \put(20,10){\usebox{\varone}}
\put(60,50){\usebox{\varone}} \put(10,60){\line(6,1){60}}
\put(10,80){\line(2,-5){20}} \dashline{5}(30,30)(10,60)
\dashline{5}(10,80)(70,70) \put(70,48){\makebox(0,0){$x_i$}}
\put(87,70){\makebox(0,0){$v_i$}}
 \put(10,42){\makebox(0,0){$x_j$}}  \put(30,8){\makebox(0,0){$x_k$}}
 \put(-7,80){\makebox(0,0){$v_j$}}  \put(-7,60){\makebox(0,0){$v'_j$}}  \put(47,30){\makebox(0,0){$v_k$}}
\end{picture}

\caption{The $\exists$snake property says that this snake pattern does not occur on value $v_i$ 
for variable $x_i$ (for any variables $x_j,x_k$ and any values $v_j,v'_j,v_k$).}

\label{fig:VEsnake}

\end{figure}

\begin{definition}  \label{def:existssnake}
A variable $x_i$ satisfies the $\exists$snake property if
$\exists v_i \in \mathcal{D}(x_i)$ such that $\forall x_j \in X
\setminus \{x_i\}$, $\forall v_j,v'_j \in \mathcal{D}(x_j)$,  $\forall x_k \in X
\setminus \{x_i,x_j\}$ $\forall v_k \in \mathcal{D}(x_k)$,
we do not have $(v_i,v_j) \notin R_{ij}$, $(v_i,v'_j) \in R_{ij}$, $(v_j,v_k)
\in R_{jk}$ and $(v'_j,v_k) \notin R_{jk}$.
\end{definition}

The proof of the following proposition can be found in 
Appendix~\ref{sec:appsnake} where we give an optimised algorithm, making use of appropriate data structures, 
to apply this variable-elimination rule until convergence. 

\begin{proposition} \label{prop:complexitysnake}
Variable eliminations by the $\exists$snake property can be applied until convergence in $O(ed^3)$ time and $O(ed^2)$ space.
\end{proposition}

As we will show in the rest of this paper, other variable-elimination rules can be found by
enriching the language of forbidden patterns by allowing arbitrary quantification and counting.
Previous work only considered quantification on values for the variable to be eliminated~\cite{ve}.

\section{Variable elimination by the DE-snake rule}  
\label{sec:DE-snake}

We show in this section that the $\exists$snake rule is subsumed by a stronger rule that we call the 
DE-snake (double-existential snake) rule.
It is again based on forbidding the snake pattern shown in Figure~\ref{fig:VEsnake} but,
compared to the $\exists$snake rule, has an existential (rather than universal) quantifier on the value $v'_j$.

\setlength{\unitlength}{1pt}
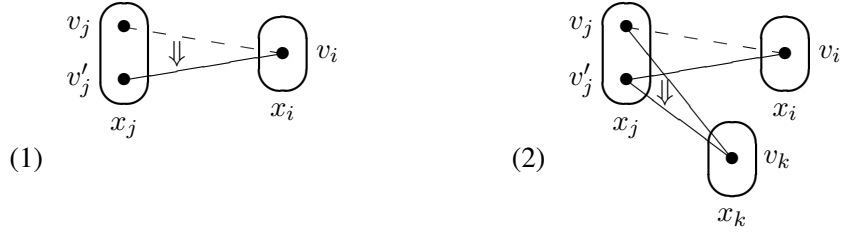
\begin{figure}
\centering

\begin{picture}(330,110)(-10,0)

\put(0,0){
\begin{picture}(140,110)(-30,0)
\put(0,50){\usebox{\vartwo}} 
\put(60,50){\usebox{\varone}} \put(10,60){\line(6,1){60}}
\dashline{5}(10,80)(70,70) \put(70,48){\makebox(0,0){$x_i$}}
\put(87,70){\makebox(0,0){$v_i$}}
 \put(10,42){\makebox(0,0){$x_j$}} 
 \put(-7,80){\makebox(0,0){$v_j$}}  \put(-7,60){\makebox(0,0){$v'_j$}}  
\put(30,70){\makebox(0,0){$\Downarrow$}}  \put(-27,30){\makebox(0,0){(1)}}  
\end{picture}}

\put(190,0){
\begin{picture}(140,110)(-30,0)
\put(0,50){\usebox{\vartwo}} \put(40,10){\usebox{\varone}}
\put(60,50){\usebox{\varone}} \put(10,60){\line(6,1){60}}
\put(10,80){\line(4,-5){40}} \put(10,60){\line(4,-3){40}}  
\dashline{5}(10,80)(70,70) \put(70,48){\makebox(0,0){$x_i$}}
\put(87,70){\makebox(0,0){$v_i$}}
 \put(10,42){\makebox(0,0){$x_j$}}  \put(50,8){\makebox(0,0){$x_k$}}
 \put(-7,80){\makebox(0,0){$v_j$}}  \put(-7,60){\makebox(0,0){$v'_j$}}  \put(67,30){\makebox(0,0){$v_k$}}
 \put(-27,30){\makebox(0,0){(2)}}   \put(24.5,55){\makebox(0,0){$\Downarrow$}} 
\end{picture}}

\end{picture}

\caption{The DE-snake property says that for some value $v_i \in \mathcal{D}(x_i)$,
for each value $v_j$ incompatible with $v_i$, there is a value $v'_j$ such that
(1) $v'_j$ is compatible with $v_i$, and 
(2) $v'_j$ is compatible with all assignments $v_k$ to a third variable $x_k$ which
are compatible with $v_j$.}

\label{fig:DEsnake}

\end{figure}

\begin{definition}  \label{def:DE-snake}
A variable $x_i$ satisfies the DE-snake property if
$\exists v_i \in \mathcal{D}(x_i)$ such that $\forall x_j \in X
\setminus \{x_i\}$, $\forall v_j \in \mathcal{D}(x_j)$ with $(v_i,v_j) \notin R_{ij}$, $\exists v'_j \in \mathcal{D}(x_j)$
such that (1) $(v'_j,v_i) \in R_{ji}$ and (2) $\forall x_k \in X \setminus \{x_i,x_j\}$, $\forall v_k \in \mathcal{D}(x_k)$,
we do not have $(v_j,v_k) \in R_{jk}$ and $(v'_j,v_k) \notin R_{jk}$.
\end{definition}

The DE-snake property is illustrated in Figure~\ref{fig:DEsnake}. The intuition behind this property is that
any solution to the instance obtained after elimination of $x_i$ can be extended to a solution to the original
instance by assigning $v_i$ to $x_i$ and changing those values $v_j$ which are incompatible with $v_i$
to some other value $v'_j$.

\begin{theorem} \label{thm:DE-snake}
The DE-snake property is a sol-var-elim condition in binary CSP instances.
\end{theorem}

\begin{altproof}
Let $I=\tuple{X,\mathcal{D},R}$ be a binary CSP instance satisfying the DE-snake property on $x_i$.
Let $I'$ be the instance obtained by eliminating variable $x_i$ from $I$.
If there is no solution to $I'$, then obviously there is none to $I$. 
Now suppose that there is a solution $s'$ to $I'$. 
We will show that $I$ has a solution $s$.
Our proof is constructive and there is an obvious polynomial-time algorithm to produce $s$ from $s'$.
Since $x_i$ satisfies the DE-snake property, $\exists v_i \in \mathcal{D}(x_i)$ such that $\forall x_j \in X
\setminus \{x_i\}$, $\forall v_j \in \mathcal{D}(x_j)$ with $(v_i,v_j) \notin R_{ij}$,  $\exists u_j(v_j) \in \mathcal{D}(x_j)$
(i.e. there exists a value $u_j$ which is a function of $v_j$) 
such that (1) $(u_j(v_j),v_i) \in R_{ji}$ and (2) $\forall x_k \in X \setminus \{x_i,x_j\}$, $\forall v_k \in \mathcal{D}(x_k)$,
we do not have $(v_j,v_k) \in R_{jk}$ and $(u_j(v_j),v_k) \notin R_{jk}$.

Let $Y$ be the set of variables $x_j \in  X \setminus \{x_i\}$ such that $(s'(x_j),v_i) \in R_{ji}$
and $\overline{Y}$ the set of variables  $x_j \in  X \setminus \{x_i\}$ such that $(s'(x_j),v_i) \notin R_{ji}$. 
For each $x_j \in Y$, set $s(x_j)$ := $s'(x_j)$. For each $x_j \in \overline{Y}$, set
$s(x_j)$ := $u_j(s'(x_j))$. Finally, set $s(x_i)$ := $v_i$.
By definition of $s$ and $u_j(s'(x_j))$, we have $(s(x_j),s(x_i)) \in R_{ji}$ for each $x_j \in  X \setminus \{x_i\}$.
For pairs of variables $x_j,x_k \in X \setminus \{x_i\}$, we need to consider three cases:
\begin{enumerate}
\item If $x_j,x_k \in Y$, then clearly $(s(x_j),s(x_k)) \in R_{jk}$ since $s'$ was a solution to $I'$.
\item If $x_j \in \overline{Y}$ and $x_k \in Y$, then setting $v_j=s'(x_j)$ and $v_k=s'(x_k)$ in the  definition of the 
DE-snake property, from condition (2) in this definition,
we must have $(u_j(s'(x_j)),s'(x_k)) \in R_{jk}$ since $(s'(x_j),s'(x_k)) \in R_{jk}$ . 
Hence $(s(x_j),s(x_k)) \in R_{jk}$
by definition of $s$.
\item If $x_j,x_k \in \overline{Y}$, then for exactly the same reason as in case 2, we again must have 
$(u_j(s'(x_j)), \\  s'(x_k)) \in R_{jk}$. In other words, by definition of $s$, $(s(x_j),s'(x_k)) \in R_{jk}$.
Now we apply again the definition of the DE-snake property but this time with the roles of the variables $x_j$,$x_k$
reversed and with $v_j=s'(x_k)$ and $v_k=s(x_j)$: we can deduce that we must have $(u_k(s'(x_k)),s(x_j)) \in R_{kj}$.
Thus, by definition of $s$,  $(s(x_k),s(x_j)) \in R_{kj}$.
\end{enumerate}
We have just shown that for all pairs of variables of $I$, $s$ satisfies the binary constraint on this pair of variables.
Hence $s$ is a solution to $I$.
\end{altproof}

The following proposition shows that the worst-case complexity of applying the DE-snake rule is no worse than
the complexity of applying the $\exists$snake rule given in Proposition~\ref{prop:complexitysnake}. 
Its proof can be found in 
Appendix~\ref{sec:appDEsnake} where we give an optimised algorithm, making use of appropriate data structures, 
to apply the DE-snake variable-elimination rule until convergence. 

\begin{proposition}
Variable eliminations by the DE-snake property can be applied until convergence in $O(ed^3)$ time and $O(ed^2)$ space. 
\end{proposition}

\section{Variable elimination by the triangle property}    \label{sec:triangle}

\setlength{\unitlength}{1pt}
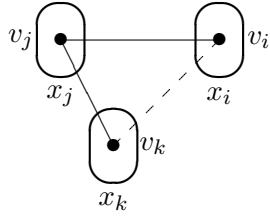
\begin{figure}
\centering

\begin{picture}(100,100)(-10,-8)
\put(0,50){\usebox{\varone}} \put(20,10){\usebox{\varone}}
\put(60,50){\usebox{\varone}} \put(10,70){\line(1,0){60}}
\dashline{4}(30,30)(70,70) \put(10,70){\line(1,-2){20}}
\put(70,49){\makebox(0,0){$x_i$}} \put(10,49){\makebox(0,0){$x_j$}}
\put(30,9){\makebox(0,0){$x_k$}} \put(85,70){\makebox(0,0){$v_i$}}
\put(-5,70){\makebox(0,0){$v_j$}} \put(45,30){\makebox(0,0){$v_k$}}
\end{picture}

\caption{
The open-triangle pattern. The variable $x_i$ can be eliminated by the triangle property if 
there is some variable $x_j \neq x_i$ such that for all $v_j \in \mathcal{D}(x_j)$,
there exists $v_i \in \mathcal{D}(x_i)$ such that $(v_j,v_i) \in R_{ji}$ and this 
open-triangle pattern does not occur on $(v_i,v_j,v_k)$ for any $v_k$.}

\label{fig:delta1}

\end{figure}

The variable-elimination rule presented in this section says that $x_i$ can be eliminated if for some variable
$x_j \neq x_i$, for all assignments $v_j \in \mathcal{D}(x_j)$ to
$y$, in $I[\tuple{x_j,v_j}]$ (the reduced instance consisting of
the set of variable-value assignments compatible with $\tuple{x_j,v_j}$) there
is an assignment $\tuple{x_i,v_i}$ compatible with all assignments to all
variables $x_k \in X \setminus \{x_i,x_j\}$. In other words, 
there is some variable $x_j \neq x_i$ such that for all $v_j \in \mathcal{D}(x_j)$,
there exists $v_i \in \mathcal{D}(x_i)$ such that $(v_j,v_i) \in R_{ji}$ and the 
open-triangle pattern shown in Figure~\ref{fig:delta1} does not occur.

\begin{definition}  \label{def:triangle}
A variable $x_i$ satisfies the  \emph{triangle property} if $\exists x_j \in X
\setminus \{x_i\}$ such that $\forall v_j \in \mathcal{D}(x_j)$, $\exists v_i
\in \mathcal{D}(x_i)$ with $(v_i,v_j) \in R_{ij}$ such that $\forall x_k \in X
\setminus \{x_i,x_j\}$, $\forall v_k \in \mathcal{D}(x_k)$, $(v_j,v_k)
\in R_{jk}$ implies that $(v_i,v_k) \in R_{ik}$.
\end{definition}

\noindent Although a variable satisfying the triangle property was originally known as `not Triangle-support\-ed'
~\cite{beyond}, 
we use the name `triangle property' in this paper for simplicity of presentation.

\begin{theorem} \label{prop:var-elim}
The triangle property is a sol-var-elim condition in binary CSP instances.
\end{theorem}

\begin{altproof}
Let $I$ be a binary CSP instance satisfying the triangle property on $x_i$. Let $s'$ be a solution
to $I'$, the instance obtained by eliminating variable $x_i$ from $I$.
We will show that $I$ has a solution $s$. Our proof is constructive and there is an obvious
polynomial-time algorithm to produce $s$ from $s'$.
Since $x_i$ satisfies the triangle property, $\exists x_j \in X
\setminus \{x_i\}$ such that $\forall v_j \in \mathcal{D}(x_j)$, $\exists
v_i(v_j) \in \mathcal{D}(x_i)$ (i.e. there exists a value $v_i$ which is a function
of $v_j$) with $(v_i(v_j),v_j) \in R_{ij}$ such that
$\forall x_k \in X \setminus \{x_i,x_j\}$, $\forall v_k \in \mathcal{D}(x_k)$
with $(v_j,v_k) \in R_{jk}$, we have $(v_i(v_j),v_k) \in R_{ik}$. Define $s$ as
follows: $s(x_m) = s'(x_m)$ ($x_m \in X \setminus \{x_i\}$) and $s(x_i) =
v_i(s'(x_j))$. The assignment $\tuple{x_i,v_i(s'(x_j))}$ is compatible with
$\tuple{x_j,s'(x_j)}$ (by definition of $v_i(s'(x_j))$) and is compatible
with all of the assignments $\tuple{x_m,s'(x_m)}$ ($x_m \in X \setminus
\{x_i\}$) again by definition of $v_i(s'(x_j))$ since $(s'(x_j),s'(x_m)) \in R_{jm}$.
Hence $s$ is a solution to $I$.

It is easily verified that this proof is valid even in the special
case $X = \{x_i,x_j\}$.
\end{altproof}

The proof of the following proposition can be found in  Appendix~\ref{sec:apptriangle}
where we give an optimised algorithm, making use of appropriate data structures, 
to apply this variable-elimination rule until convergence. 

\begin{proposition}
Variable eliminations by the triangle property can be applied until convergence in $O(end^3)$ time and $O(end^2)$ space.
\end{proposition}

\section{From broken triangles to broken polyhedra}  \label{sec:from}

The broken-triangle property is a property of the microstructure of
instances of the binary CSP (Constraint Satisfaction Problem) which
when satisfied allows either value merging~\cite{OnBT-AIJ}, variable elimination or
the definition of a tractable class~\cite{Cooper10:btp}. In this section, we generalise the notion of
broken triangle to broken polyhedron, which allows us to define
rules for variable elimination parameterised by the dimension $k$ of the polyhedron.

We begin by recalling the definition of the broken-triangle property (BTP)~\cite{Cooper10:btp}.
\begin{definition}
\label{defn:BTP} Let $I=\tuple{X,\mathcal{D},R}$ be a binary CSP instance.
A pair of values $v'_k,v''_k \in \mathcal{D}(x_k)$ satisfies BTP if for each pair of
variables $(x_i,x_j)$ (with $i,j \neq k$), $\forall v_i \in \mathcal{D}(x_i)$, $\forall v_j \in \mathcal{D}(x_j)$, if
\begin{itemize}
\item [$\bullet$] $(v_i,v_j) \in R_{ij}$,
\item [$\bullet$] $(v_i,v'_k) \in R_{ik}$ and
\item [$\bullet$] $(v_j,v''_k) \in R_{jk}$,
\end{itemize}
then
\begin{itemize}
\item [$\bullet$] $(v_i,v''_k) \in R_{ik}$ or
\item [$\bullet$] $(v_j,v'_k) \in R_{jk}$.
\end{itemize}
A variable $x_k$ satisfies BTP if each pair of values
of $\mathcal{D}(x_k)$ satisfies BTP. If $I$ is
equipped with an order $<$ on its variables, then $I$ satisfies BTP 
for the variable order $<$ if
each variable $x_k$ satisfies BTP in the sub-instance
of $I$ restricted to the variables $x_i$ such that $x_i \leq x_k$.
\end{definition}

\begin{figure}[ht]
\centerline{\scalebox{1}[1]{\includegraphics{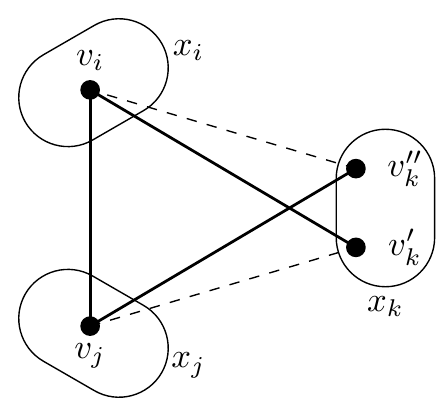}}}
\caption{A broken triangle $(v'_k,v_i,v_j,v''_k)$.}
\label{btp}
\end{figure}

If $(v_i,v_j) \in R_{ij}$,$(v_i,v'_k) \in R_{ik}$,
$(v_j,v''_k) \in R_{jk}$, $(v_i,v''_k) \notin R_{ik}$ and $(v_j,v'_k) \notin R_{jk}$ (as shown in
Figure \ref{btp} with solid/broken lines joining compatible/incompatible values), then the 
quadruple ($v'_k,v_i,v_j,v''_k$) constitutes a \emph{broken triangle} on $x_k$.
$I$ satisfies the BTP on $x_k$ if no broken triangles occur on $x_k$.

Any pair of values $v'_k,v''_k$ that satisfy BTP can be merged without changing the satisfiability of the instance~\cite{OnBT-AIJ},
where the new merged value is compatible with all values compatible with at least one of the two old values $v'_k,v''_k$.
Furthermore, in an arc-consistent instance any variable that satisfies BTP can be eliminated 
without changing the satisfiability of the instance~\cite{Cooper10:btp}.
As a direct consequence of this, if an arc-consistent instance $I$ satisfies BTP
for some variable ordering, then $I$ can be solved in polynomial time by successive elimination of all variables: 
moreover viewed as a decision problem, $I$ can, in fact, be solved by arc consistency~\cite{Cooper10:btp}
without knowledge of the variable ordering for which BTP holds.

Examples of the BTP var-elim rule include a variable $x_m$ which is only constrained by
one other variable in an arc-consistent instance or a 
variable $x_m$ with a domain of size at most two in a path-consistent instance~\cite{Cooper10:btp}. 
In this paper we consider generalisations of BTP which allow the definition of stronger variable-elimination rules.

The presence of some broken triangles on a given variable does not
preclude value-merging or variable-elimination (while leaving satisfiability invariant): new solutions 
will not be introduced if the broken triangles lack support on some set of 
other variables~\cite{wBTP,DBLP:journals/constraints/Mouelhi18,wadySETN}.
Unfortunately, the search for lack-of-support variables for each broken triangle may render such techniques 
prohibitively expensive in terms of time complexity (since, in the worst case, the number of broken triangles 
is $\Theta(n^3d^4)$). Other generalisations of BTP
require levels of consistency, such as strong path consistency, which may change positive edges
into negative edges; this has the disadvantage of possibly introducing new 
broken triangles besides the extra memory required to store new 
binary constraints~\cite{kbtp,DBLP:journals/jetai/Naanaa13}.

We now generalise the notion of broken triangle to broken polyhedron. 
In Section~\ref{sec:existential} we show that this notion 
can be used to define variable elimination rules that are stronger than BTP.
A broken triangle is a broken polyhedron of dimension 2. We now define a
broken $k$-dimensional polyhedron for $k \geq 2$.

\begin{definition}
\emph{A broken $k$-dimensional polyhedron} on $x_m$ consists of
\begin{itemize}
\item  [$\bullet$] a consistent assignment $\langle v_1,\ldots,v_k \rangle$ to distinct variables $\langle
x_{i_1},\ldots,x_{i_k} \rangle$ (where each $x_{i_j}$
($j=1,\ldots,k$) is distinct from $x_m$),
\item  [$\bullet$] $k$ distinct values $u_1,\ldots,u_k \in D_m$,
\end{itemize}
such that
\begin{itemize}
\item  [$\bullet$] $\forall j \in \{1,\ldots,k\}$, $(v_j,u_j) \notin R_{i_jm}$,
\item  [$\bullet$] $\forall h,j \in \{1,\ldots,k\}$, if $h \neq j$ then $(v_h,u_j) \in R_{i_jm}$,
\end{itemize}
The assignment $\langle v_1,\ldots,v_k \rangle$ to variables $\langle
x_{i_1},\ldots,x_{i_k} \rangle$ is known as the \emph{base} of the
broken polyhedron, and each assignment $\langle x_m,u_j \rangle$
($j=1,\ldots,k$) is an \emph{apex}. Any edge between a base point and an apex
is a \emph{side} of the  broken tetrahedron.
\end{definition}

\thicklines
\setlength{\unitlength}{2pt}
\begin{figure}
\centerline{\scalebox{1}[1]{\includegraphics{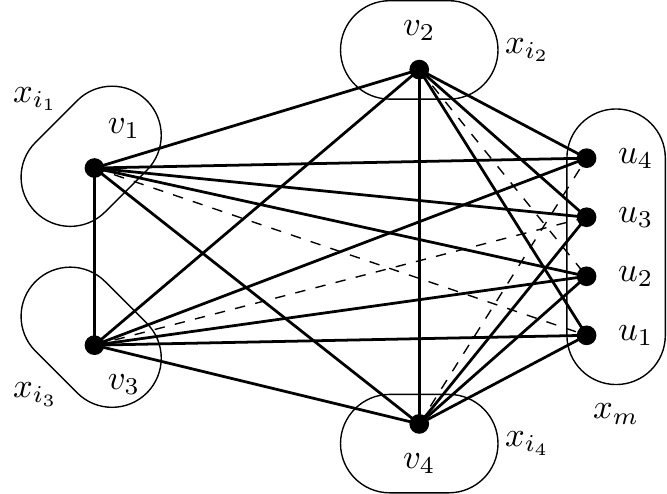}}}


\caption{A broken 4-dimensional polyhedron on variable $x_m$.} \label{fig:poly4}
\end{figure}
\setlength{\unitlength}{1pt}

A broken 4-dimensional polyhedron is shown in Figure~\ref{fig:poly4}. A broken triangle
(Figure~\ref{btp}) is a broken 2-dimensional polyhedra.
In the following sections we show that broken $k$-dimensional polyhedra allow us to define novel 
variable-elimination rules and tractable classes.

\section{First-order rules for variable-elimination by broken polyhedra} \label{sec:existential}

The broken-triangle property (BTP) has been generalised to the $\forall\exists$BTP rule
for variable elimination which allows us to eliminate more variables~\cite{beyond}
than BTP.  Eliminating a variable satisfying the $\forall\exists$BTP rule is strictly stronger
than the BTP  rule. This is demonstrated by the fact that $\forall\exists$BTP, but not BTP,
subsumes the rule that allows us to eliminate a variable $x_m$
when an assignment to $x_m$ is compatible with all assignments to all
other variables. Another generic example is when all occurrences of
the broken-triangle pattern 
on variable $x_m$ occur on pairs of values $v_m,v'_m \in S \subset \mathcal{D}(x_m)$ and each
assignment $v_i$ to each other variable $x_i \neq x_m$ has a support at
$x_m$ in $\mathcal{D}(x_m) \setminus S$.

We first given the definition of 
the $\forall\exists$ broken-triangle property, in order to generalise it to $k$ dimensions.

\begin{definition} \label{def:AEbtp}
A binary CSP instance satisfies the \emph{$\forall\exists$ broken-triangle property} on variable $x_m$ if for all
$i_1 \neq m$, for all $v_1 \in \mathcal{D}(x_{i_1})$, there exists $v_m \in \mathcal{D}(x_m)$ such that 
\begin{enumerate}
\item  $\langle v_1,v_m \rangle$ is a consistent assignment to variables $\langle x_{i_1},x_m \rangle$,  and 
\item for all $i_2 \notin \{i_1,m\}$, for all $v_2 \in \mathcal{D}(x_{i_2})$,
there is no broken triangle on $x_m$ with base the
assignment $\langle v_1,v_2 \rangle$ to variables $\langle x_{i_1},x_{i_2} \rangle$ and with 
an apex $\langle x_m, v_m \rangle$.
\end{enumerate}
\end{definition}

We now generalise the $\forall\exists$BTP rule for variable elimination to the case
of broken polyhedra of any dimension $k \geq 2$. When $k=2$ the following
definition coincides with Definition~\ref{def:AEbtp} of the $\forall\exists$BTP rule.

\begin{definition} \label{def:bpp}
A binary CSP instance satisfies the \emph{$\forall\exists$ broken
$k$-dimensional polyhedron property} on variable $x_m$ if for all
distinct $i_1,\ldots,i_{k-1} \neq m$, for all 
consistent assignments $\langle v_1,\ldots,v_{k-1} \rangle$ to variables
$\langle x_{i_1},\ldots,x_{i_{k-1}} \rangle$, there exists $v_m \in
\mathcal{D}(x_m)$ such that 
\begin{enumerate}
\item  $\langle v_1,\ldots,v_{k-1},v_m \rangle$ is a
consistent assignment to variables $\langle
x_{i_1},\ldots,x_{i_{k-1}},x_m \rangle$,  and 
\item for all $i_k \notin \{i_1,\ldots,i_{k-1},m\}$, for all $v_k \in \mathcal{D}(x_{i_k})$,
there is no broken $k$-dimensional polyhedron on $x_m$ with base the
assignment $\langle v_1,\ldots,v_k \rangle$ to variables $\langle
x_{i_1},\ldots,x_{i_k} \rangle$ and with an apex $\langle x_m, v_m
\rangle$.
\end{enumerate}
\end{definition}

The first condition of Definition~\ref{def:bpp} is a $k$-consistency condition on variable $x_m$ with respect to
all other variables~\cite{lecoutre}. The second condition guarantees (as we will show below) that this $k$-consistency condition is
sufficient for any consistent assignment to the variables $X \setminus \{x_m\}$ to be extendible to
a consistent assignment to all variables.

\noindent {\bf Notation:} Given a binary CSP instance $I$ on variables 
$X = \{x_1,\ldots,x_n\}$, we denote by $I_{-m}$ the sub-instance of $I$ on variables 
$X \setminus \{x_m\}$. Similarly, $I_{-jm}$ denotes the sub-instance
of $I$ on variables $X \setminus \{x_j,x_m\}$.

We can observe that if $I$ satisfies the $\forall\exists$ broken $k$-dimensional polyhedron property on $x_m$
then $I_{-j}$ also satisfies the $\forall\exists$ broken $k$-dimensional polyhedron property on $x_m$ for any $j \neq m$.

\begin{theorem}  \label{thm:poly}
The $\forall\exists$ broken $k$-dimensional polyhedron property is a sol-var-elim condition in
binary CSP instances $I$ with at least $k$ variables.
\end{theorem}

\begin{altproof}
Let $I$ be a binary CSP instance $I$ on $n \geq k$ variables which satisfies the
$\forall\exists$ broken $k$-dimensional polyhedron property on variable $x_m$.
It is sufficient to show that any solution $s$ to $I_{-m}$ can be extended to a solution
to $I$. We will show this by induction on $n$. That a solution for $I$ can be generated
in polynomial time will follow immediately since $s$ does not need to be modified, just extended
by one of the at most $d$ possible values for $x_m$.

If $n=k$, then the fact that any solution to $I_{-m}$ can be extended to a solution to $I$
follows directly from the definition of the $\forall\exists$ broken $k$-dimensional polyhedron property.
So, to complete the proof by induction, we suppose that 
any solution to $I_{-m}$ can be extended to a solution to $I$ holds for instances $I$ with $n=p \geq k$
variables and we will show that this also holds for instances with $p+1$ variables.

For notational convenience and without loss of generality, we can assume that $m=p+1$.
Let $s := \langle v_1,\ldots,v_p \rangle$ be a solution to $I_{-m}$. To complete the proof, it suffices to show that $s$
can be extended to a solution to $I$. For each $j=1,\ldots,p$,
consider the $p$-variable instance $I_{-j}$. Clearly, $s_{j} :=  \langle v_1,\ldots,v_{j-1},v_{j+1},\ldots,v_p \rangle$
is a solution to $I_{-jm}$. Since $I_{-j}$ has $p$ variables, by our inductive hypothesis, each 
$s_{j}$ ($j=1,\ldots,p$) can be extended to a solution $t_j = \langle v_1,\ldots,v_{j-1},v_{j+1},\ldots,v_p, u_j \rangle$
to $I_{-j}$. Note that $t_j$ assigns $u_j$ to $x_m$.
Suppose that for some $j \in \{1,\ldots,p\}$, $(v_j,u_j) \in R_{jm}$. Then 
$\langle v_1,\ldots,v_p,u_j \rangle$ is a solution to $I$ and we are done.
So, we only need consider the case in which $\forall j \in \{1,\ldots,p\}$, $(v_j,u_j) \notin R_{jm}$.
Note that the values $u_j$ ($j=1,\ldots,p$) must all be distinct.

Consider $\langle v_1,\ldots,v_{k-1} \rangle$; this is a consistent assignment to
variables $\langle x_1,\ldots,x_{k-1} \rangle$. Thus, since $I$ satisfies the
$\forall\exists$ broken $k$-dimensional polyhedron property on variable $x_m$, there exists $v_m \in
\mathcal{D}(x_m)$ such that (1) $\langle v_1,\ldots,v_{k-1},v_m \rangle$ is a
consistent assignment to the variables $\langle
x_1,\ldots,x_{k-1}, \\ x_m \rangle$  and (2) for all $h \in \{k,\ldots,p\}$, 
there is no broken $k$-dimensional polyhedron on $x_m$ with base the
assignment $\langle v_1,\ldots,v_{k-1},v_h \rangle$ to variables $\langle
x_1,\ldots,x_{k-1},x_h \rangle$ and with an apex $\langle x_m, v_m \rangle$.
Note that $v_h$ is the value assigned by $s$ to $x_h$. Observe that
$\langle v_1,\ldots,v_{k-1},v_h \rangle$ is a consistent assignment to variables $\langle
x_1,\ldots,x_{k-1},x_h \rangle$ and that $u_1, \ldots$, $u_{k-1}, v_m$ are distinct values
(since $\forall j \in \{1,\ldots,p\}$, $(v_j,u_j) \notin R_{jm}$ but $(v_j,v_m) \in R_{jm}$,
and we have already seen above that the $u_j$ ($j=1,\ldots,p$) are distinct).
Furthermore, by the definition of the $u_j$, we have $(v_h,u_j) \in R_{hm}$ for $j=1,\ldots,k-1$,
and $(v_\ell,u_j) \in R_{\ell m}$ if and only if $\ell \neq j$ for $\ell,j \in \{1,\ldots,k-1\}$.
By the definition of $v_m$ (condition (1) above), we have $(v_j,v_m) \in R_{jm}$ for $j=1,\ldots,k-1$.
Since there is no broken $k$-dimensional polyhedron on $x_m$ with base the
assignment $\langle v_1,\ldots,v_{k-1},v_h \rangle$ to variables $\langle
x_1,\ldots,x_{k-1},x_h \rangle$ and with apex $\langle x_m, v_m \rangle$, we must have
$(v_h,v_m) \in R_{hm}$. Since this is true for all $h \in \{k,\ldots,p\}$, 
it follows that $\langle v_1,\ldots,v_p,v_m \rangle$ is a solution to $I$.
This completes the proof by induction. 
\end{altproof}

\section{Faster variable-elimination based on broken polyhedra}   \label{faster_ve}

The $\forall\exists$ broken $k$-dimensional polyhedron property is interesting from a theoretical point of view.
However, from a practical point of view, the time complexity of detecting whether variables can be
eliminated is likely to be prohibitive. Indeed, for $k=3$, a naive exhaustive search for broken tetrahedra
(i.e. broken 3-dimensional polyhedra) has time complexity $\Theta(n^4d^6)$.
Various intermediate rules of varying strength and time complexity exist between $\forall\exists$BTP
and the $\forall\exists$ broken tetrahedron property; we choose to concentrate on rules which
can be tested in the same worst-case time complexity as $\forall\exists$BTP but which are strictly stronger.

\thicklines
\setlength{\unitlength}{2pt}
\begin{figure}
\centerline{\scalebox{1}[1]{\includegraphics{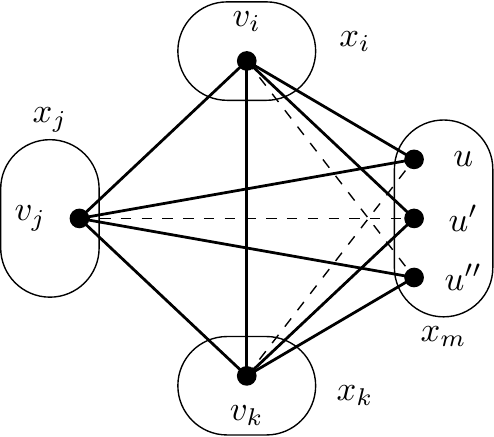}}}

\caption{A broken tetrahedron.} \label{fig:tetra}
\end{figure}

First, observe that a broken tetrahedron, as shown in Figure~\ref{fig:tetra}
with base the assignment $\langle v_i,v_j, \\v_k \rangle$ to variables $\langle x_i,x_j,x_k \rangle$
and apexes $u,u',u''\in \mathcal{D}(x_m)$, contains three broken triangles on $x_m$: $(u',v_i,v_j,u'')$,
$(u,v_i,v_k,u'')$ and $(u,v_j,v_k,u')$. Thus, the incompatible pairs (shown as broken lines in
Figure~\ref{fig:tetra}) $(v_i,u'')$ and $(v_j,u')$ each occur in (at least) two broken triangles, while
the compatible pairs (shown as solid lines in the figure) $(v_i,v_j)$, $(v_i,u)$ and $(v_j,u)$ each occur in
(at least) one broken triangle.

\begin{definition} \label{def:bt_degree}
Let $I$ be a binary CSP instance with $v_i \in \mathcal{D}(x_i)$ and $u \in \mathcal{D}(x_m)$.
The broken-triangle degree (BT degree) of the pair of assignments $\langle v_i,u \rangle$ to
$\langle x_i,x_m \rangle$ is the number of distinct variables $x_j$ ($j \neq i,m$) such that 
$\exists v_j \in \mathcal{D}(x_j)$, $\exists u' \in \mathcal{D}(x_m)$ such that there is a broken triangle
with base the assignment $\langle v_i,v_j \rangle$ to $\langle x_i,x_j \rangle$
and apexes $u,u' \in \mathcal{D}(x_m)$. 
\end{definition}
For example, if $I$ is exactly the instance shown in Figure~\ref{fig:tetra}, then 
the BT degree of $(v_i,u'')$ and of $(v_j,u')$ is two, and the BT degree of $(v_i,u)$ and of $(v_j,u)$ is one.
Although $(v_i,v_j)$ is the base of a broken triangle, 
its BT degree is zero according to Definition~\ref{def:bt_degree}, since neither $v_i$ nor $v_j$
is the apex of a broken triangle.
Note that any edge, whether positive or negative, which links the base to an apex of a
broken triangle has BT degree at least one.

\begin{definition}
A consistent assignment $\langle v_i,v_j \rangle$ to variables $\langle x_i,x_j \rangle$ is \emph{3-safe}
on variable $x_m$ if for all broken triangles $(u',v_i,v_j,u'')$, the BT degree of $(v_i,u'')$ is one
or the BT degree of $(v_j,u')$ is one.
\end{definition}
Note that $\langle v_i,v_j \rangle$ is trivially 3-safe on $x_m$ if there is no broken triangle on $x_m$
whose base is $\langle v_i,v_j \rangle$.

We can now define a new variable-elimination property.

\begin{definition}  \label{def:BTdegree}
A binary CSP instance satisfies the \emph{BT-degree property} on variable $x_m$ if for all
distinct $i,j \neq m$, for all $v_i \in \mathcal{D}(x_i), v_j \in \mathcal{D}(x_j)$ such that $\langle
v_i,v_j \rangle$ is a consistent assignment to variables
$\langle x_i,x_j \rangle$, there exists $v_m \in \mathcal{D}(x_m)$ such that 
\begin{enumerate}
\item  $\langle v_i,v_j,v_m \rangle$ is a
consistent assignment to variables $\langle x_i,x_j,x_m \rangle$, and 
\item \emph{either} $(v_i,v_j)$ is 3-safe on $x_m$
\emph{or} $(v_i,v_m)$ has BT degree zero
\emph{or} $(v_j,v_m)$ has BT degree zero.
\end{enumerate}
\end{definition}

The first condition in Definition~\ref{def:BTdegree} guarantees path consistency on variable $x_m$
with respect to all other pairs of variables~\cite{lecoutre}, whereas the second condition
guarantees the absence of a broken 3-dimensional polyhedron with base points $v_i,v_j$ and apex $v_m$.

\begin{theorem}  \label{thm:limited}
The BT-degree property is a sol-var-elim condition in binary CSP instances $I$ with at least 3 variables.
\end{theorem}

\begin{altproof}
It suffices, by Theorem~\ref{thm:poly}, to show that a binary CSP instance $I$
that satisfies the BT-degree property on $x_m$ necessarily satisfies the 
$\forall\exists$ broken 3-dimensional polyhedron property on $x_m$. But this is immediate
by the discussion above since 
\begin{enumerate}
\item pairs of assignments 
which are 3-safe on $x_m$ cannot be part of the base of a broken tetrahedron on $x_m$,
by the remark before Definition~\ref{def:bt_degree},
\item pairs of assignments which have BT degree zero cannot be the side 
(i.e. edge between a base point and an apex) of a broken tetrahedron.   \hfill 
\end{enumerate} 
\end{altproof}

\par
\begin{figure}[ht]
  \centerline{(a) \scalebox{1}[1]{\includegraphics{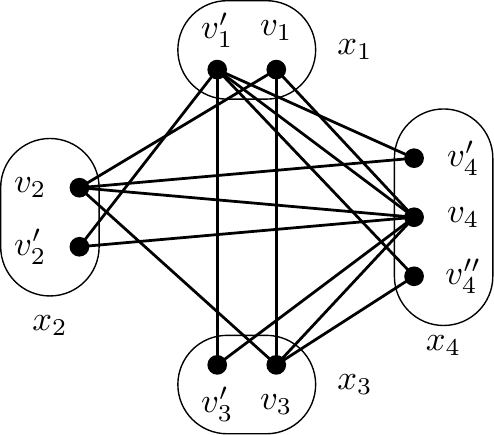}} \hspace{1.3cm} (b)
   \scalebox{1}[1]{\includegraphics{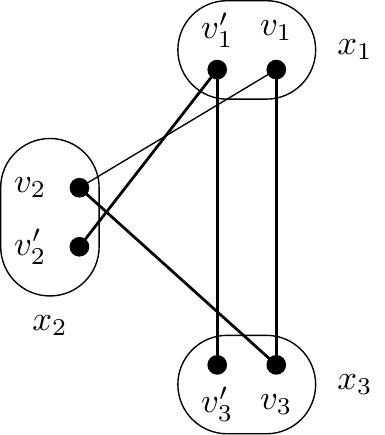}}}
\caption{(a) $x_1$ cannot be eliminated (by the BT-degree property) in this instance $I$, 
(b) $x_1$ can be eliminated in this instance $I'$ which results after elimination of $x_4$ from $I$.}
\label{fig:var-elim}
\end{figure}

 
Variable eliminations by the  BT-degree property may propagate. 
In the instance $I$ of Figure~\ref{fig:var-elim}(a) (where in this figure 
 pairs of values not joined by a line are assumed to be incompatible), the variables $x_1,x_2,x_3$ cannot be
eliminated by Theorem~\ref{thm:limited} since the consistent assignments $(v_3,v_4'')$,
$(v_1',v_4')$ and $(v_1',v_4'')$ have no support, respectively, at $x_1$, $x_2$ and $x_3$.
However, we can eliminate $x_4$ by Theorem~\ref{thm:limited}, since  
$v_4$ is a support at $x_4$ of any consistent assignment $(v_i,v_j)$ to any pair of 
variables $(x_i,x_j)$ ($1 \!  \leq \!  i \!  < \!  j \!  \leq \! 3$) 
and in each case $(v_i,v_4)$ has BT degree 0
(since $v_4$ is consistent with all assignments to all other variables). 
Eliminating $x_4$ then produces an instance $I'$ (shown in Figure~\ref{fig:var-elim}(b))
in which $x_1$ can now be eliminated
by Theorem~\ref{thm:limited} since the only consistent assignment $(v_2,v_3)$ to $(x_2,x_3)$
can be extended to the consistent assignment $(v_2,v_3,v_1)$ to $(x_2,x_3,x_1)$ and $(v_2,v_1)$ has BT degree 0.

The proof of the following proposition can be found in Appendix~\ref{sec:appBTdegree}
where we give an optimised algorithm, using appropriate data structures, 
to apply this variable-elimination rule until convergence. 

\begin{proposition}
The BT-degree variable-elimination property 
can be applied until convergence in $O(end^3)$ time and $O(end^2)$ space.
\end{proposition}

We can compare this with the $\forall\exists$BTP property (Definition~\ref{def:bpp} with $k=2$) 
which can also be applied until convergence
in $O(end^3)$ time and $O(end^2)$ space, as shown in Appendix~\ref{sec:appBTP}.
Eliminating variables by the broken triangle property, which is subsumed by the $\forall\exists$BTP property,
also has time complexity $O(end^3)$~\cite{Cooper10:btp}. 
Thus the BT-degree property is comparable with the $\forall\exists$BTP property 
in terms of computational complexity. The following proposition shows that it is at least as powerful
in terms of number of variables eliminated.

\begin{proposition}  \label{prop:btprules}
If a binary CSP instance satisfies $\forall\exists$BTP
on variable $x_m$ then it satisfies the BT-degree property on variable $x_m$.
\end{proposition}

\begin{altproof}
The $\forall\exists$BTP property says that for all $i \neq m$, for all $v_i \in \mathcal{D}(x_i)$, 
$\exists d(v_i) \in \mathcal{D}(x_m)$ (i.e. there exists a value $d$ which is a function of $v_i$) 
such that $(v_i,d(v_i)) \in R_{im}$ and the BT degree of $(v_i,d(v_i))$ is zero.
To show that this implies the BT-degree property on variable $x_m$,
consider any pair of distinct variables $x_i,x_j$ ($i,j \neq m$) and any values
$v_i \in \mathcal{D}(x_i)$, $v_j \in \mathcal{D}(x_j)$ such that $(v_i,v_j) \in R_{ij}$. We need to show that
there exists $v_m \in \mathcal{D}(x_m)$ such that 
\begin{enumerate}
\item $(v_i,v_m) \in R_{im}$ and $(v_j,v_m) \in R_{jm}$, and
\item $(v_i,v_j)$ is 3-safe on $x_m$ or the BT degree of $(v_i,v_m)$ is zero or the BT degree of $(v_j,v_m)$ is zero.
\end{enumerate}
If $(v_j,d(v_i)) \in R_{jm}$, then $v_m = d(v_i)$ satisfies the first condition and the BT degree of $(v_i,v_m)$ is zero.
If $(v_i,d(v_j)) \in R_{im}$, then $v_m = d(v_j)$ satisfies the first condition and the BT degree of $(v_j,v_m)$ is zero.
It is not possible to have both $(v_j,d(v_i)) \notin R_{jm}$ and $(v_i,d(v_j)) \notin R_{im}$ since, in this case,
$(d(v_i),v_i,v_j,d(v_j))$ would be a broken triangle, contradicting the fact that the BT degree of $(v_i,d(v_i))$ is zero.
We can therefore conclude that in all cases, the BT-degree property is satisfied
on variable $x_m$.
\end{altproof}

\thicklines \setlength{\unitlength}{1.5pt}
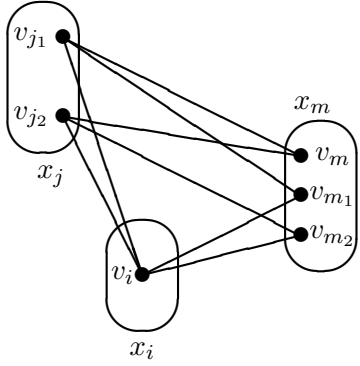
\begin{figure}
\centering

\begin{picture}(100,100)(0,0)
\put(10,60){\usebox{\vartwobigbis}}
\put(30,10){\usebox{\varonebig}}
\put(70,30){\usebox{\varthreebigbis}}
\put(20,90){\line(2,-1){60}} \put(20,90){\line(3,-2){60}} \put(20,90){\line(1,-3){20}} 
\put(20,70){\line(6,-1){60}} \put(20,70){\line(2,-1){60}} \put(20,70){\line(1,-2){20}}  
\put(40,30){\line(2,1){40}} \put(40,30){\line(4,1){40}} 
\put(40,11){\makebox(0,0){$x_i$}}  \put(17,55){\makebox(0,0){$x_j$}}  
\put(83,73){\makebox(0,0){$x_m$}} 
\put(35,30){\makebox(0,0){$v_i$}}
\put(88,60){\makebox(0,0){$v_m$}}
\put(88,50){\makebox(0,0){$v_{m_1}$}} 
\put(88,40){\makebox(0,0){$v_{m_2}$}}
\put(12,90){\makebox(0,0){$v_{j_1}$}} 
\put(12,70){\makebox(0,0){$v_{j_2}$}}
\end{picture}

\caption{An instances in which $x_m$ can be eliminated by the BT-degree property but
not by the $\forall\exists$BTP property.}
\label{fig:strictlysubsumes}
\end{figure}

In fact, the BT-degree property strictly subsumes the $\forall\exists$BTP property,
as illustrated by the instance in Figure~\ref{fig:strictlysubsumes}. In this instance, $x_m$ cannot be
eliminated by $\forall\exists$BTP since for $v_i \in \mathcal{D}(x_i)$, there is no $v \in \mathcal{D}(x_m)$
with $(v_i,v) \in R_{im}$ and such that the BT degree of $(v_i,v)$ is zero (because of the broken triangles
$(v_{m_1},v_i,v_{j_2},v_m)$ and $(v_{m_2},v_i,v_{j_1},v_m)$). On the other hand, $x_m$ can be eliminated by the 
BT-degree property since for both of the assignments $(v_i,v_{j_k})$ $(k=1,2)$
to $(x_i,x_j)$, there is a value $v_{m_k} \in \mathcal{D}(x_m)$ with $(v_i,v_{m_k}) \in R_{im}$, 
$(v_{j_k},v_{m_k}) \in R_{jm}$ and such that the BT degree of $(v_{j_k},v_{m_k})$ is zero.

\section{Theoretical comparison between different variable-elimination rules}  \label{sec:theory}
The notion of rank introduced by Naanaa~\cite{DBLP:journals/jetai/Naanaa13} 
is closely related to the absence of broken $k$-dimensional polyhedra. Indeed, 
in a binary CSP instance, a variable $x_m$ has rank $k-1$
if there is no broken $k$-dimensional polyhedron on $x_m$. Naanaa 
showed that a variable $x_m$ with rank at most $k-1$ in a binary CSP instance which is directional
strong $k$-consistent~\cite{lecoutre} (according to an order which places $x_m$ last)
can be eliminated while leaving the satisfiability of the instance invariant.
This is subsumed by the $\forall\exists$ broken $k$-dimensional polyhedron property (Definition~\ref{def:bpp}) 
since the latter does not require the absence of all broken $k$-dimensional polyhedra~\cite{DBLP:conf/dagstuhl/CooperZ17}.

Following an orthogonal approach, it has recently 
been shown that  singleton arc consistency
~\cite{lecoutre}
solves instances that do not contain a pattern (known as $Q1$) made up of 
a subset of the edges of a broken tetrahedron~\cite{stacs18}.

Another family of variable-elimination rules is $m$-fBTP~\cite{DBLP:journals/constraints/Mouelhi18}, 
for $m \geq 1$, which extends BTP by allowing broken triangles that do not have a support at some subset of variables
of size $m$.

\begin{definition}  \label{def:1fBTP}
Let $I=\tuple{X,\mathcal{D},R}$ be a binary CSP instance.
 A pair of values $v'_k,v''_k \in \mathcal{D}(x_k)$ satisfies $1$-fBTP if for each broken triangle 
($v'_k,v_i, v_j,v''_k$) with $ v_i \in \mathcal{D}(x_i)$, $v_j \in \mathcal{D}(x_j)$, 
there is at least one variable $x_{\ell} \in X \setminus \{ x_i,x_j,x_k\}$  such that $\forall  v_{\ell} \in \mathcal{D}(x_{\ell})$,   
if $(v_i,v_{\ell}) \in R_{i \ell}$ then $(v_j,v_{\ell}) \notin R_{j \ell}$. 
In this case, we say that $x_{\ell}$ is a  support variable for the broken triangle ($v'_k,v_i,v_j,v''_k$). 
A variable $x_k \in X$ satisfies $1$-fBTP if each pair of values $v'_k,v''_k \in \mathcal{D}(x_k)$  satisfies $1$-fBTP. 
\end{definition}

The variable $x_m$ of the instance shown in Figure \ref{fig:tetra} does not satisfy $1$-fBTP because there is no support 
variable for the broken triangle ($u,v_i,v_k,u''$). The concept of support variable can be extended to a set of $m$ 
variables to obtain the definition of $m$-fBTP~\cite{DBLP:journals/constraints/Mouelhi18}.

\thicklines \setlength{\unitlength}{1.5pt}

\begin{figure}[t]
\centerline{%
\begin{tabular}{c@{\hspace{3cm}} c}
\scalebox{1}[1]{\includegraphics{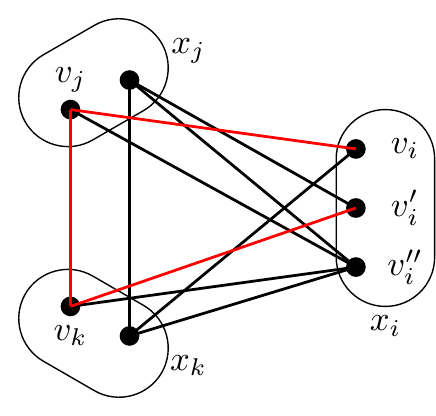}} & \scalebox{1}[1]{\includegraphics{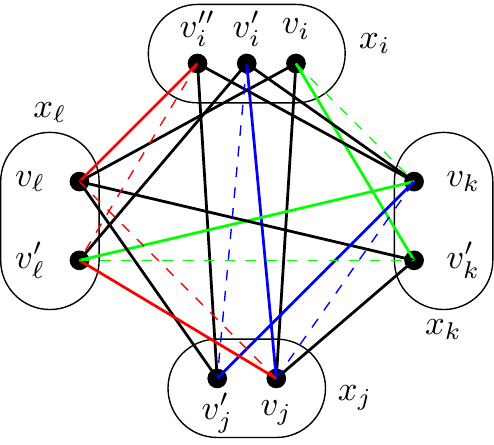}}\\
 & \\
(a) & (b) \\
 & \\
 & \\
\scalebox{1}[1]{\includegraphics{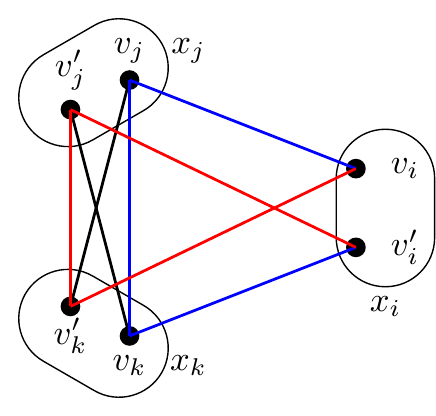}} & \scalebox{1}[1]{\includegraphics{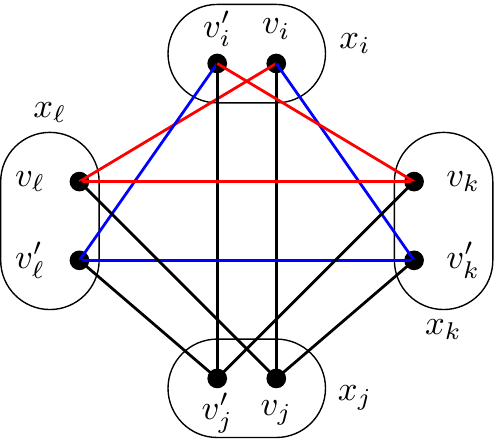}}\\
 & \\
(c) & (d) \\
\end{tabular}}
\caption{Examples of instances to demonstrate the incomparability of the different variable-elimination rules.}
\label{fig:examples}
\end{figure}

In the rest of the paper, we use the notation ($\exists$, DE-)snake in the following sense: 
a statement is true for ($\exists$, DE-)snake property
if it is true for both the $\exists$snake property and the DE-snake property.
We now compare theoretically the four variable-elimination rules: the
($\exists$, DE-)snake property, the triangle property, $\forall\exists$BTP and $1$-fBTP.
Two variable-elimination rules are incomparable if neither is subsumed by the other.

\begin{proposition}  \label{prop:incomparable}
The following four variable-elimination rules are all pairwise incomparable: the
($\exists$, DE-)snake property, the triangle property, $\forall\exists$BTP and $1$-fBTP.
\end{proposition}
\begin{altproof}
Figure~\ref{fig:examples} shows four binary CSP instances. In this figure, compatible  
values are joined by lines, and hence incompatibility is represented by the absence of a line.
In the instance shown in Figure~\ref{fig:examples}(a),
variable $x_i$ can be eliminated by the $\exists$snake property, the triangle property or $\forall\exists$BTP, but not by $1$-fBTP.
Indeed, a variable $x_i$ for which there exists $v''_i \in \mathcal{D}(x_i)$ compatible with all
values for all other variables (as is the case for $x_i$ in Figure~\ref{fig:examples}(a))
can always be eliminated by the $\exists$snake (and also the DE-snake) property, 
the triangle property or $\forall\exists$BTP, but not necessarily by $1$-fBTP.
In this example, there is a broken triangle ($v'_i$,$v_k$,$v_j$,$v_i$), shown in red, 
and (trivially) no other variable
on which this broken triangle does not have a support, so $x_i$ cannot be eliminated by $1$-fBTP.

In the instance shown in Figure~\ref{fig:examples}(b), there is no broken triangle on
variable $x_i$, so it can be eliminated by $\forall\exists$BTP or $1$-fBTP, but not by the 
($\exists$, DE-)snake property nor the triangle property.
It is easily verified that the snake pattern (Figure~\ref{fig:VEsnake}) occurs on each value 
$u \in \mathcal{D}(x_i)$ for some other variable $x_m$ and for each positive edge $uv$
with $v \in \mathcal{D}(x_m)$ (the snake patterns are  
represented by three different colours with the negative edges of the pattern shown as dashed lines).

In the instance shown in Figure~\ref{fig:examples}(c), variable $x_i$ can be eliminated by the $\exists$snake property 
(and hence also the DE-snake property which subsumes the $\exists$snake property), 
but none of the triangle property, $\forall\exists$BTP or $1$-fBTP.
The $\exists$snake property is satisfied since the snake pattern  (Figure~\ref{fig:VEsnake}) 
does not occur on $v_i \in \mathcal{D}(x_i)$. The triangle property is not satisfied on $x_i$ since
the open-triangle pattern shown in Figure~\ref{fig:delta1} occurs on $v_j \in \mathcal{D}(x_j)$ and on $v_k \in \mathcal{D}(x_k)$.  
The broken triangles ($v_i$,$v_j$,$v_k$,$v'_i$) and ($v_i$,$v'_k$,$v'_j$,$v'_i$), respectively 
shown in blue and red, prevent elimination of $x_i$ by $\forall\exists$BTP or $1$-fBTP.

In the instance shown in Figure~\ref{fig:examples}(d), 
variable $x_i$ can be eliminated by the triangle property or $1$-fBTP, but not by $\forall\exists$BTP nor 
the ($\exists$, DE-)snake property.
The triangle property is satisfied on $x_i$ since, 
for all $v \in \mathcal{D}(x_j)$ no open-triangle pattern (illustrated in Figure~\ref{fig:delta1}) occurs on $v$. 
The broken triangles ($v_i$,$v_\ell$,$v_k$,$v'_i$) and ($v'_i$,$v'_\ell$,$v'_k$,$v_i$),  respectively  
shown in red and blue, prevent elimination of $x_i$ by $\forall\exists$BTP; 
but these broken triangles have a support variable $x_j$ which allows $x_i$ to be eliminated by $1$-fBTP.

It can easily be verified that all combinations are covered by these four examples: for any two distinct rules, 
rule1 and rule2, among the ($\exists$, DE-)snake property, 
the triangle property, $\forall\exists$BTP and $1$-fBTP, there is
an example instance in Figure~\ref{fig:examples} such that rule1 eliminates $x_i$ but rule2 does not.
\end{altproof}

Proposition~\ref{prop:btprules} tells us that the BT-degree property subsumes
$\forall\exists$BTP. On the other hand, as we now show, it is incomparable with the three other properties.

\begin{proposition}
The BT-degree property is incomparable with each of the following 
variable elimination properties:  the ($\exists$, DE-)snake property, the triangle property and $1$-fBTP.
\end{proposition}

\begin{altproof}
The proof is identical to the proof of Proposition~\ref{prop:incomparable}, since the instances
shown in Figure~\ref{fig:examples} in which variable $x_i$ satisfies the BT-degree property 
are exactly the same instances in which $x_i$ satisfies $\forall\exists$BTP. 
In particular, $x_i$ does not satisfy the BT-degree property in Figure~\ref{fig:examples}(c) 
(respectively Figure~\ref{fig:examples}(d)) since $(v_j,v_k)$ (respectively $(v_\ell,v_k)$) 
cannot be extended to a consistent assignment for $x_i$.
Proposition~\ref{prop:btprules} tells us that $x_i$ can be eliminated by BT-degree property 
in the instances in Figure~\ref{fig:examples}(a) and Figure~\ref{fig:examples}(b) since it can be eliminated by the weaker
property $\forall\exists$BTP.
\end{altproof}

Another important generic example is the case in which $x_i$ is constrained by a single other variable $x_j$.
Such a variable $x_i$ can always be eliminated (remembering that eliminating a variable means first
deleting from $\mathcal{D}(x_j)$ all values $v_j$ with no support at $x_i$).
In this case, variable $x_i$ can be eliminated by any of  the triangle property, $\forall\exists$BTP or $1$-fBTP,
but not necessarily by the ($\exists$, DE-)snake property.

\section{Experimental results}  \label{sec:experiments}
In this section, we study the practical interest of some variable elimination rules, namely the BT-degree property, 
the $\exists$snake rule, the DE-snake rule and the triangle property.
For each rule, we assess its ability to eliminate variables and its impact on solving efficiency.
First, we describe the experimental protocol we used.

\subsection{Experimental protocol}
We considered all the binary instances from the 2008 International CP Competition\footnote{http://www.cril.univ-artois.fr/CPAI08} and we discarded those whose inconsistency is detected by
enforcing arc-consistency. By so doing, we obtained a benchmark of 3,557 CSP instances.
These instances have between 3 and 5,000 variables whose initial domains have between 2 and 10,000 values. The number of constraints varies from 3 to 124,750. Constraints are defined in extension
or in intension. For example, among these instances, we can find frequency allocation problems or graph colouring instances.

Regarding the variable elimination algorithms, for each rule, we first enforced arc-consistency and then 
eliminated those variables having a singleton domain. Note that,  in an arc-consistent
instance, singleton-domain variables would be eliminated by all four of the rules we are comparing. 
Then we applied the considered elimination rule until convergence (i.e. a fixpoint is reached at which no more
eliminations are possible by this rule). To do this, we consider a set of variables containing all the candidates for elimination.
Initially, this set contains all the variables having a non-singleton domain. 
For each candidate variable, we check whether the rule applies. 
If so, the variable is eliminated and all its neighbours are added to the candidate set. 
Two variables are neighbours if they constraint each other.
Checking whether the rule applies is performed as described in Section \ref{faster_ve} for the BT-degree property, or by a naive approach for the $\exists$snake rule,
the DE-snake rule and the triangle property.
We used more naive algorithms than those which are optimised 
for worst-case time complexity. The algorithms in the Appendix (which use data structures to reduce
worst-case time complexity) are given for their theoretical rather than practical interest,
since our first concern in these experimental trials was to estimate and compare the number of variable eliminations
that can be achieved by each rule.
In order to solve CSP instances, we used the state-of-the-art algorithm MAC+RST+NG \cite{DBLP:journals/jsat/LecoutreSTV07}.
We exploit a geometric restart policy based on the number of allowed backtracks. Initially, the number of allowed backtracks is set to 100 and the increasing factor to 1.1.
The search was guided by the dom/wdeg variable heuristic \cite{DBLP:conf/ecai/BoussemartHLS04}.
All the algorithms are written in C++ in our own library.

The experiments were performed on Dell PowerEdge M620 blade servers with Intel Xeon E5-2609 2.4 GHz processors.
We allotted 30 minutes and 16 GB of memory for each elimination process while, 
for the solving process, the timeout was set to one hour.

\subsection{Ability to eliminate variables}
\begin{figure}[tb]
\centerline{
\scalebox{0.8}[0.8]{\includegraphics{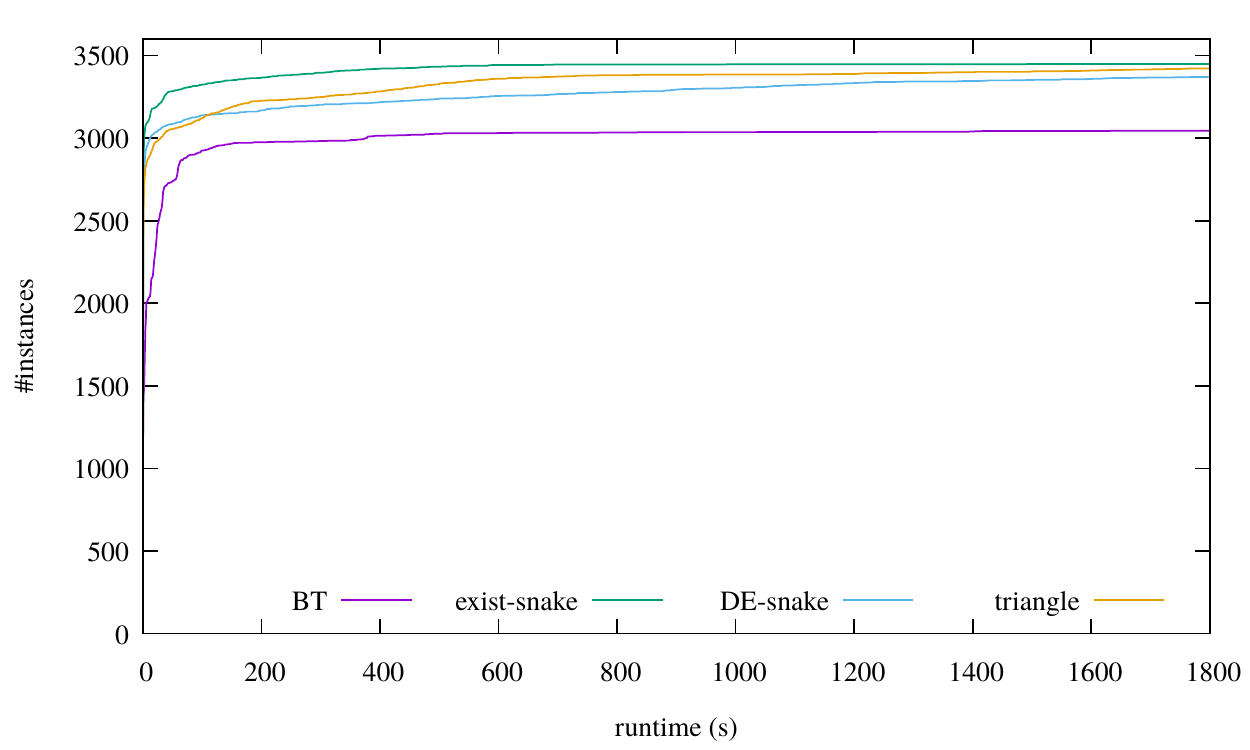}}}
\caption{Cumulative number of processed instances with respect to elapsed time.}
\label{runtime_elim}
\end{figure}

In this part, we assess the practical behaviour of the considered elimination rules and their ability to eliminate variables.
Table \ref{tab_elim} provides the number of instances for which the elimination process finishes, runs out of time or memory or is able to eliminate at least one variable.
Clearly, the ($\exists$,DE-)snake rules and the triangle property are able to process more instances than the 
BT-degree property. Such a result was foreseeable since the algorithm we used for the 
latter property has worse time and space complexities than the two others.
These complexities also explain why the elimination process based on the BT-degree property runs out of time or memory.
However, despite this, the BT-degree property still succeeds in processing about 58\% of the instances.
Moreover, as we can see in Figure \ref{runtime_elim}, which depicts the cumulative number of processed instances 
with respect to elapsed time, it processes about 90\% of the treated instances
in less than one minute. This ratio is close to the one obtained for the triangle property, while the $\exists$snake rule (respectively the DE-snake rule) can be applied for more than 95\% (resp. 92\%) of the treated instances in the same time.
An obvious conclusion we can draw from Figure~\ref{runtime_elim} is that a much shorter time-out would not have 
greatly reduced the number of variable eliminations by any of the rules. Indeed, the fact that the curves all flatten out
fairly quickly indicates that more efficient algorithms would not have detected a significantly larger number of
variable eliminations within any given time-out period.

\begin{table}[tb]
\caption{Number of instances for which the elimination process finishes, runs out of time or memory or is able to eliminate at least one variable.}
\centerline{
\begin{tabular}{|l|r|r|r|r|}
\cline{2-5}
\multicolumn{1}{c|}{} & \multicolumn{1}{c|}{BT-degree} & \multicolumn{1}{c|}{$\exists$snake}& \multicolumn{1}{c|}{DE-snake} & \multicolumn{1}{c|}{triangle}\\
\hline
\# Processed         & 2,056 &	 3,449 & 3,371	& 3,420\\
\# Timeout           &   513 &    108  &   184	&	137\\
\# Memory-out        &   988 &      0  &     0  &     0\\
\# Elim. instances   &   507 &	   786 &   836  & 1,313\\
\hline
\end{tabular}}

\label{tab_elim}
\end{table}

\begin{figure}[t]
\centerline{%
\begin{tabular}{c c}
\scalebox{0.6}[0.6]{\includegraphics{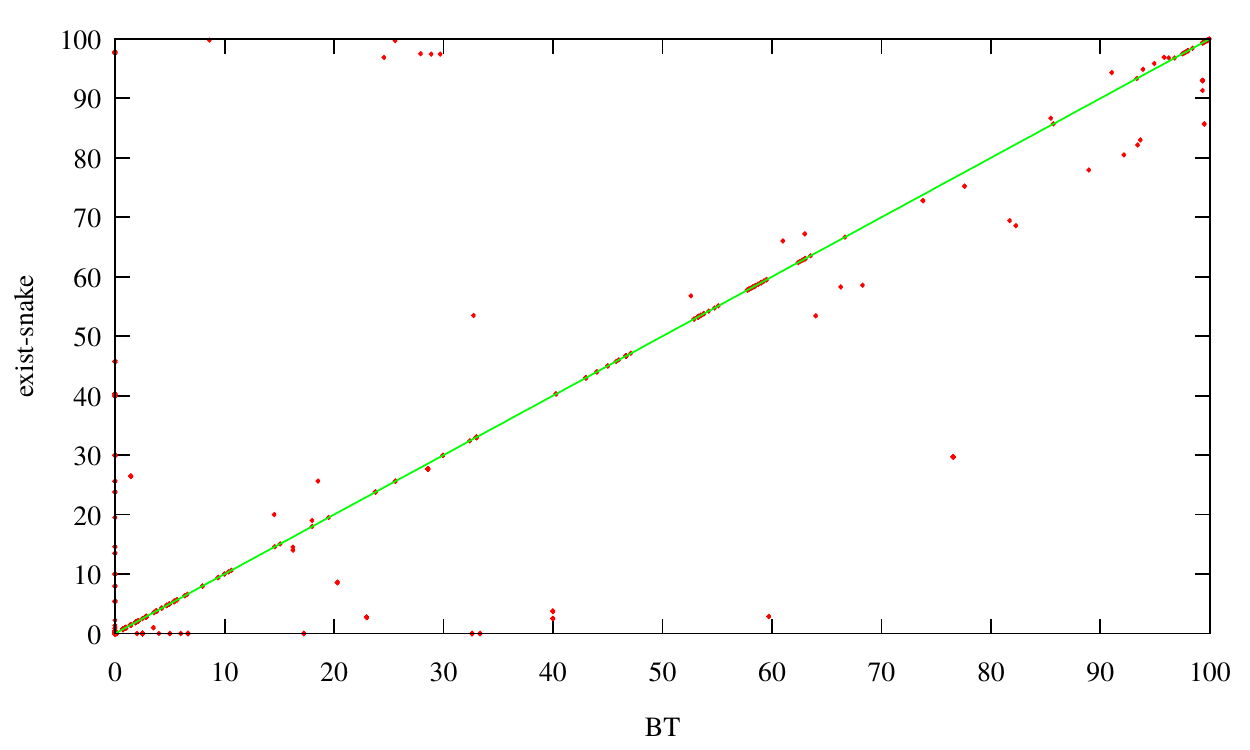}} & \scalebox{0.6}[0.6]{\includegraphics{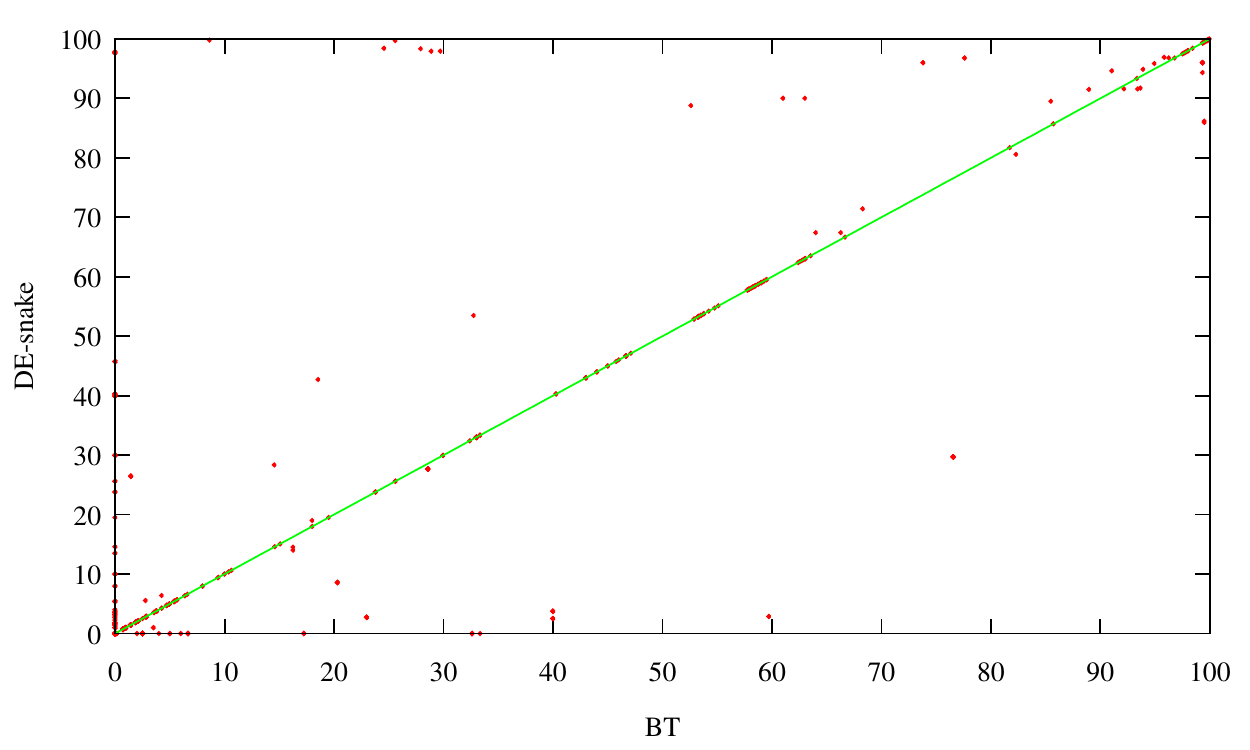}}\\
\scalebox{0.6}[0.6]{\includegraphics{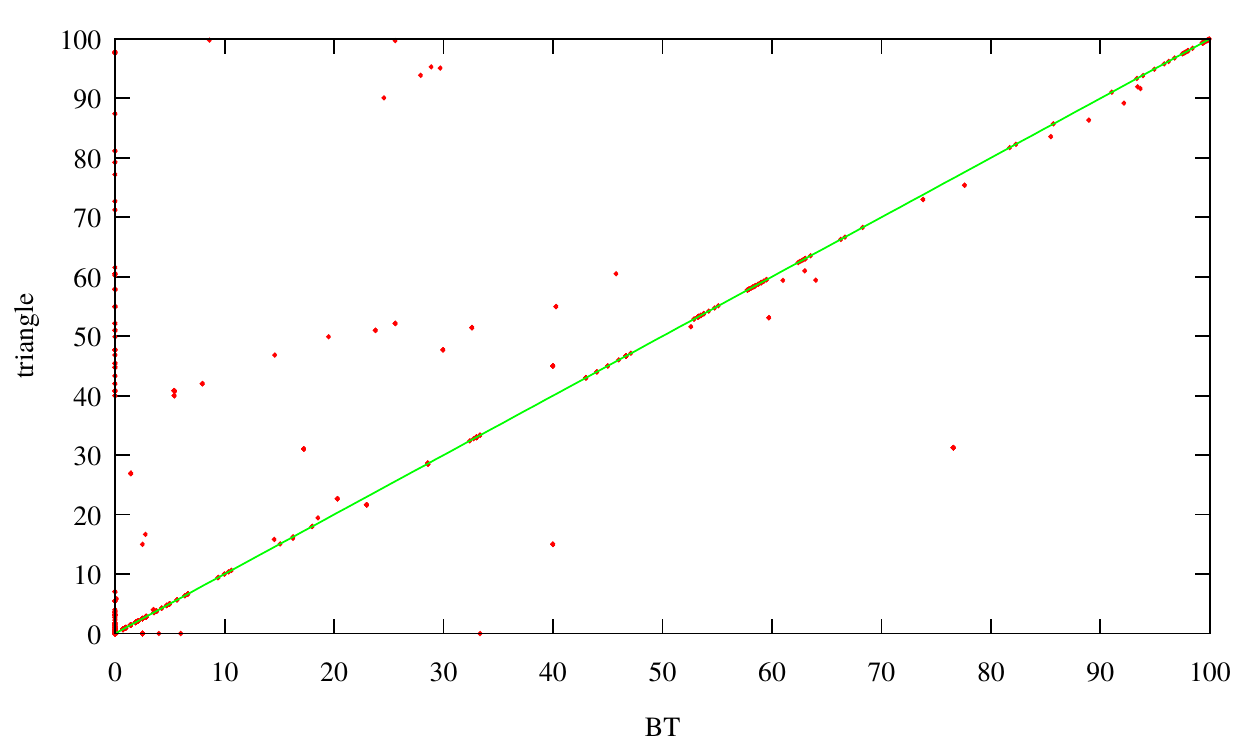}} & \scalebox{0.6}[0.6]{\includegraphics{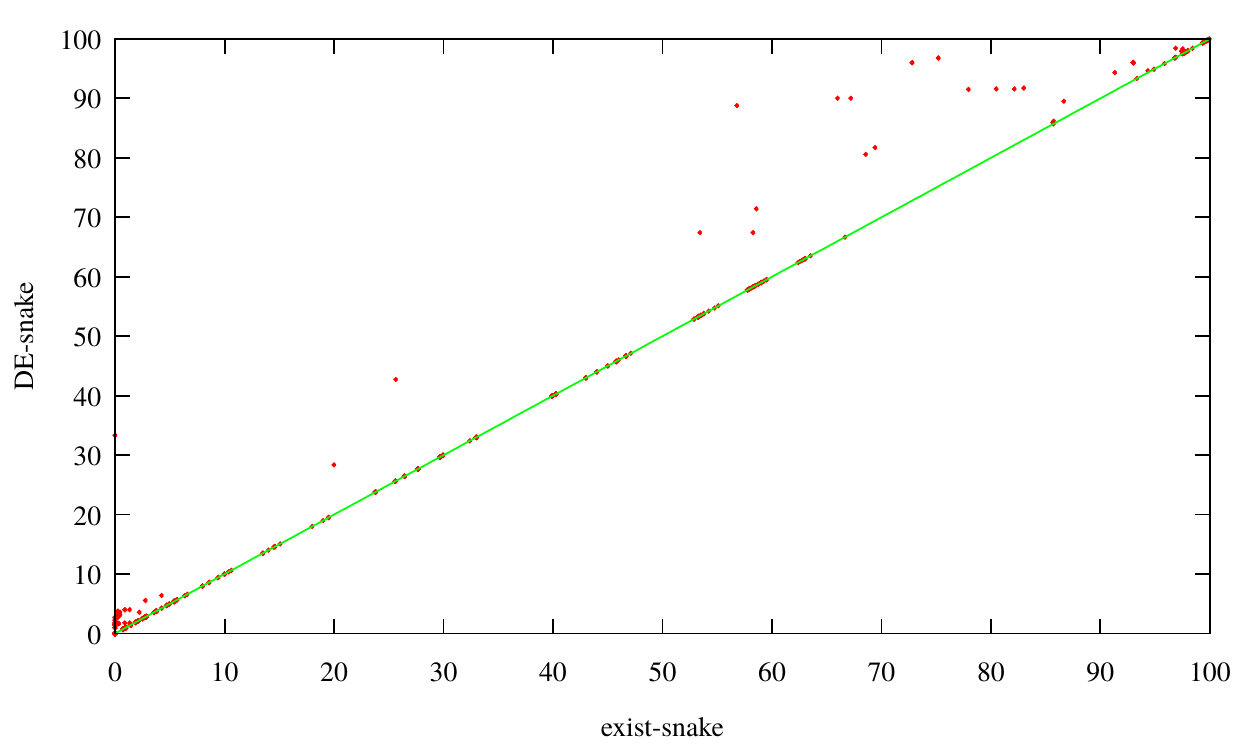}}\\
\scalebox{0.6}[0.6]{\includegraphics{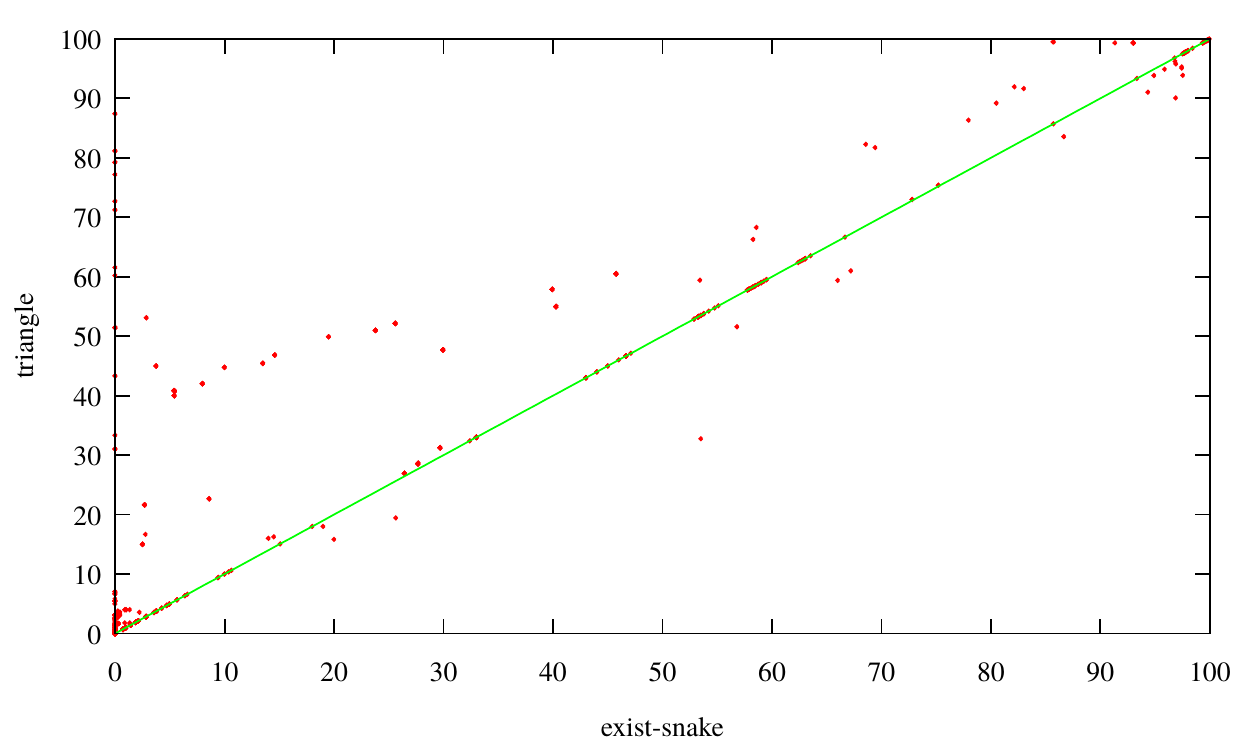}} & \scalebox{0.6}[0.6]{\includegraphics{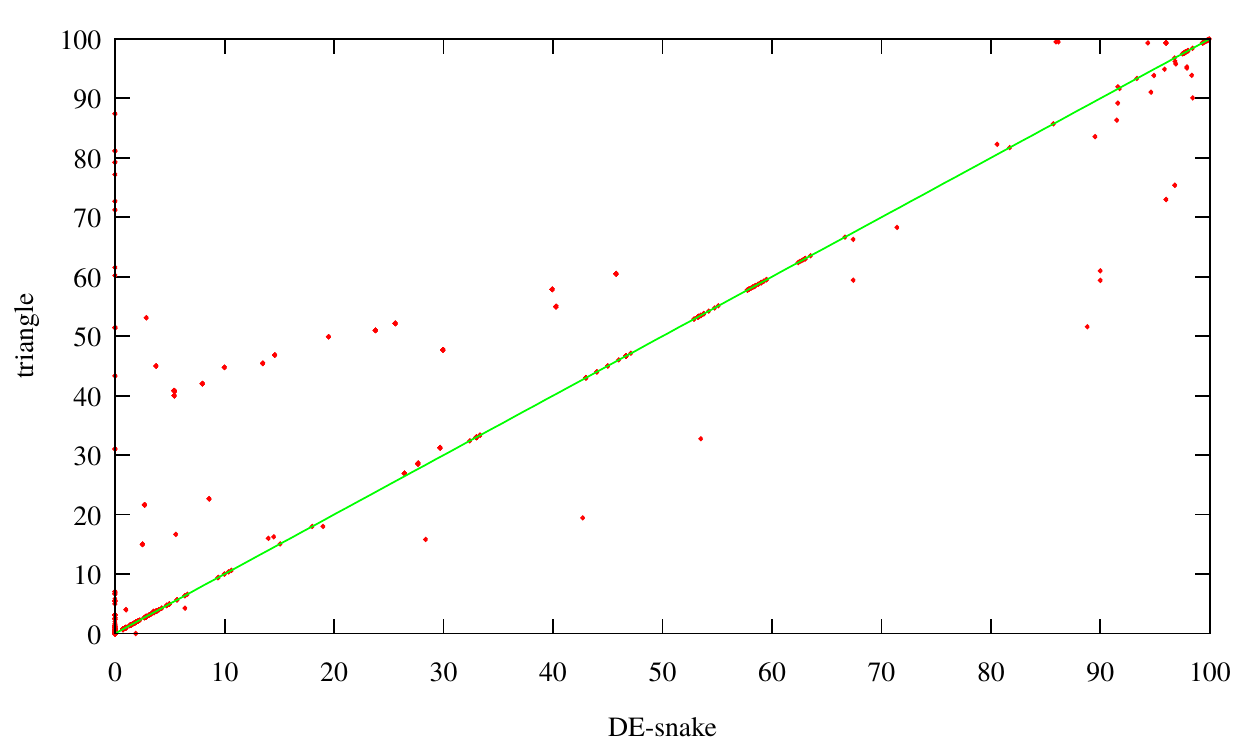}}\\
\end{tabular}}
\caption{Comparison of the percentage of eliminated variables for each instance and for each pair of elimination rules.}
\label{percentage_elim}
\end{figure}

Now, if we investigate the ability to eliminate variables, the triangle property is the most interesting rule in the sense that 
it is able to eliminate variables in more instances. The triangle rule eliminates at least one variable in 1,313 instances.
Figure \ref{percentage_elim} provides a comparison of the percentage of eliminated variables 
for each instance and for each pair of elimination rules.
First, we can remark that these comparisons are consistent with the theoretical results we provided previously. 
In particular, they illustrate the fact that the DE-snake rule subsumes the 
$\exists$snake rule and any other pair of elimination rules are incomparable.
For example, if we compare the ($\exists$,DE-)snake rules and the BT-degree property, we clearly see that they 
are incomparable since there exist instances for which some variables are eliminated by the first elimination rules and not by the second and conversely. 
Moreover, we can note that the number of the instances for which the ($\exists$,DE-)snake rules eliminate more variables is close to the respective number for the BT-degree property.
In contrast, if we compare the triangle property with the BT-degree property, the triangle 
property turns out to be more effective for eliminating variables, even if the two rules are incomparable.
Indeed, there exist only a few instances for which the BT-degree property eliminates more variables than the triangle property.
Then, if we compare the triangle property with the $\exists$snake rule, we can draw the same conclusion.
Finally, the comparison between the triangle property and the DE-snake rule seems to be more less clear.
However, we can note that for a significant number of instances, the triangle property is able to eliminate 
variables while the DE-snake rule eliminates none.
So, it turns out that the triangle property appears to be the best elimination rule with respect to 
the ability to eliminate variables.

\begin{figure}[tbh]
\centerline{
\scalebox{0.75}[0.75]{\includegraphics{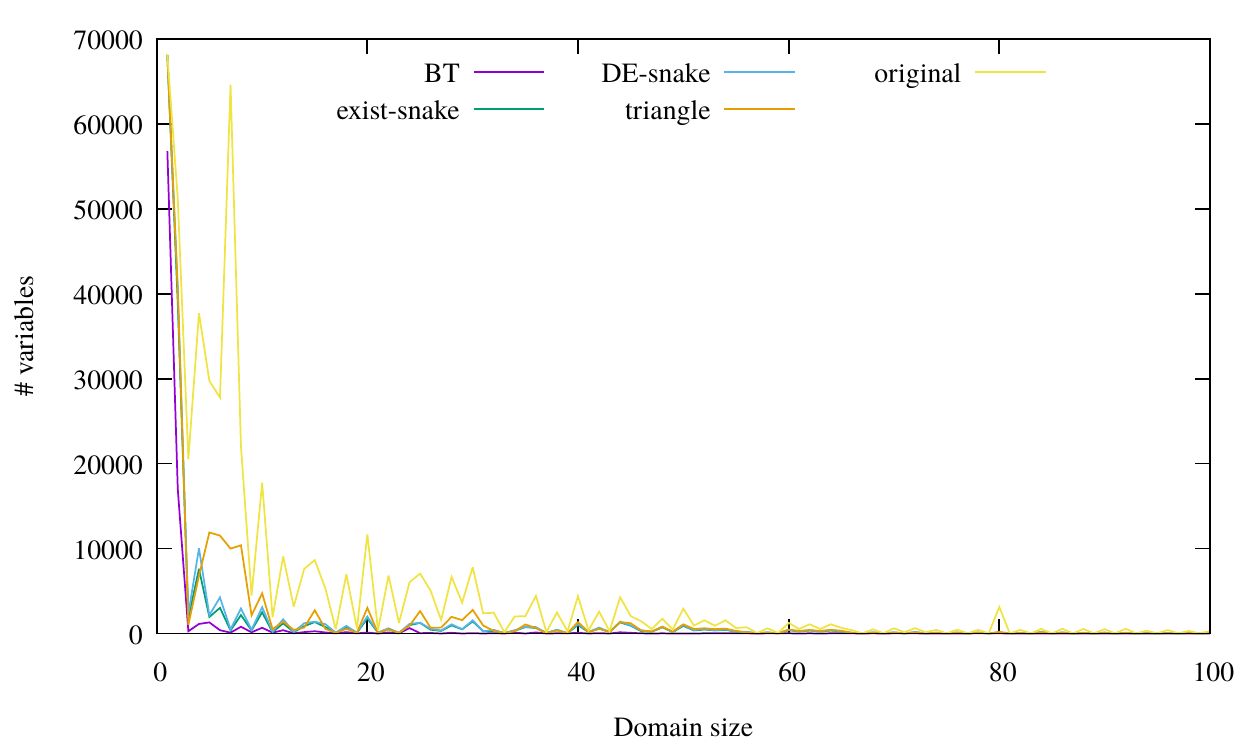}}}
\caption{Number of eliminated variables having a domain of a given size for each elimination rule and number of such variables in the original instances.}
\label{domain}
\end{figure}

\begin{figure}[tbh]
\centerline{
\scalebox{0.75}[0.75]{\includegraphics{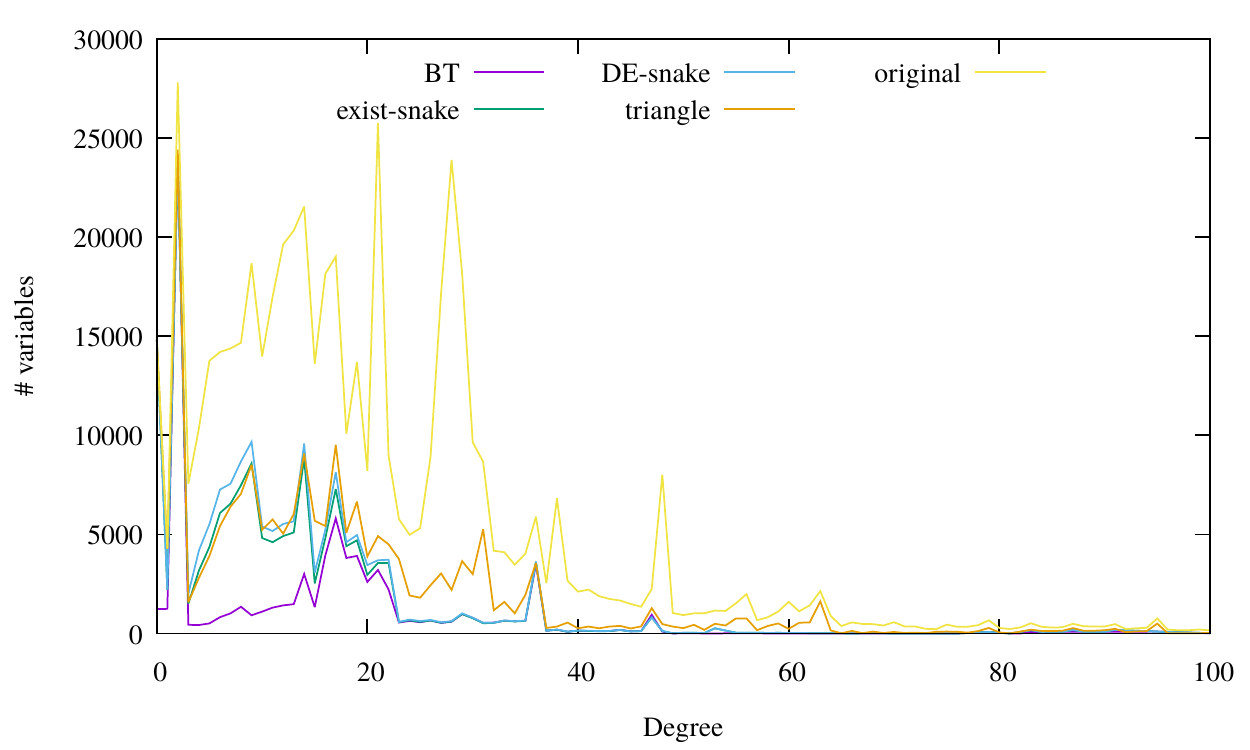}}}
\caption{Number of eliminated variables having a given degree for each elimination rule and number of such variables in the original instances.}
\label{degree}
\end{figure}

Further experiments would be needed to identify which type of variables are more likely to be eliminated by each rule. 
If it turns out that such variables have some easily identifiable characteristic, such as
a small number of neighbours or a small domain, this will help us target specific variables. It may even turn out
that these variables are exactly those for which testing the variable-elimination rules is less costly 
in computational resources.
Our preliminary investigations we made in this direction seem to show that this is the case.
Indeed, they establish that the eliminated variables often have a small domain or a small number of neighbours as shown in 
Figures \ref{domain} and \ref{degree}.
The two figures compare respectively the number of eliminated variables having a given domain size or a given degree (i.e. number of neighbours) to the corresponding number in the original 
instances. We focus our study on variables having a domain size or a degree at most 100. Above, the number of eliminated variables is negligible (about one percent in the best cases).
We can observe that in our experiments the BT-degree property did not eliminate all variables with a singleton domain. 
This is simply explained by the instances for which the elimination process runs out of memory.

\subsection{Impact on solving efficiency}
\begin{figure}[tb]
\centerline{
\scalebox{0.75}[0.75]{\includegraphics{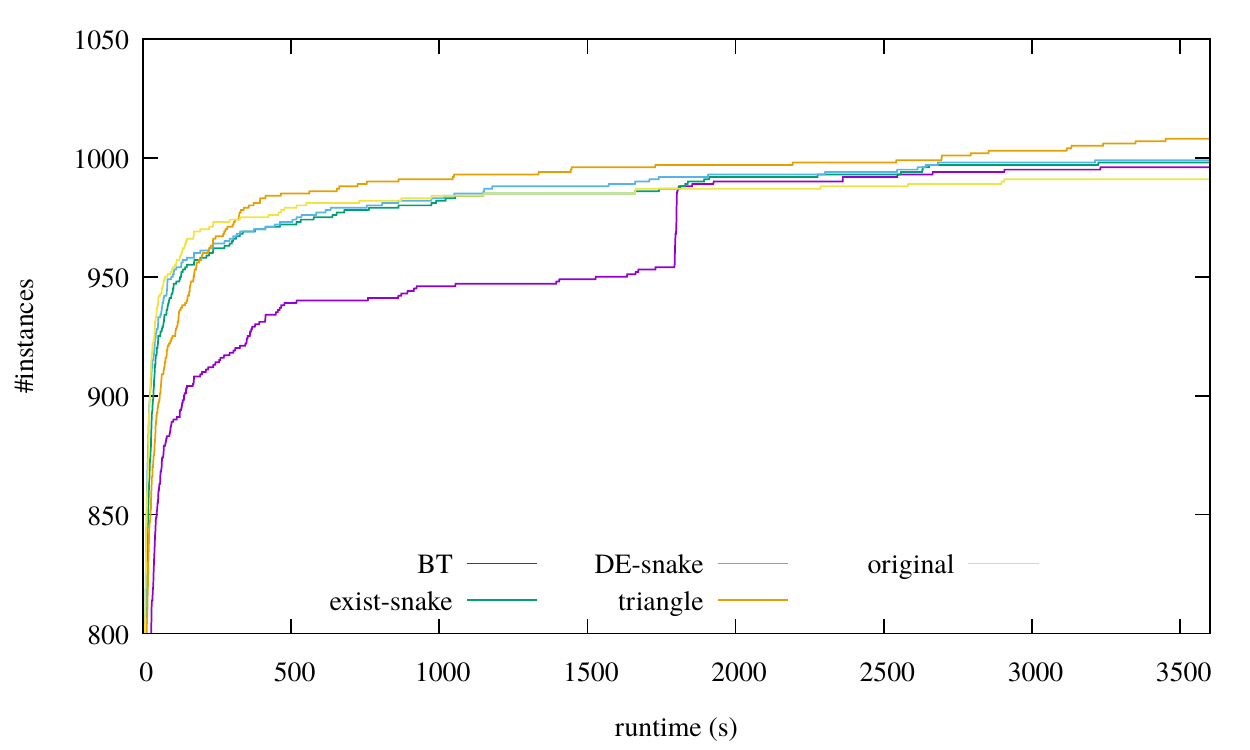}}}
\caption{Cumulative number of instances solved  by MAC+RST+NG after eliminating some variables or by considering the original instances with respect to elapsed time.}
\label{runtime_solving}
\end{figure}

\begin{figure}[tb]
\centerline{
\scalebox{0.75}[0.75]{\includegraphics{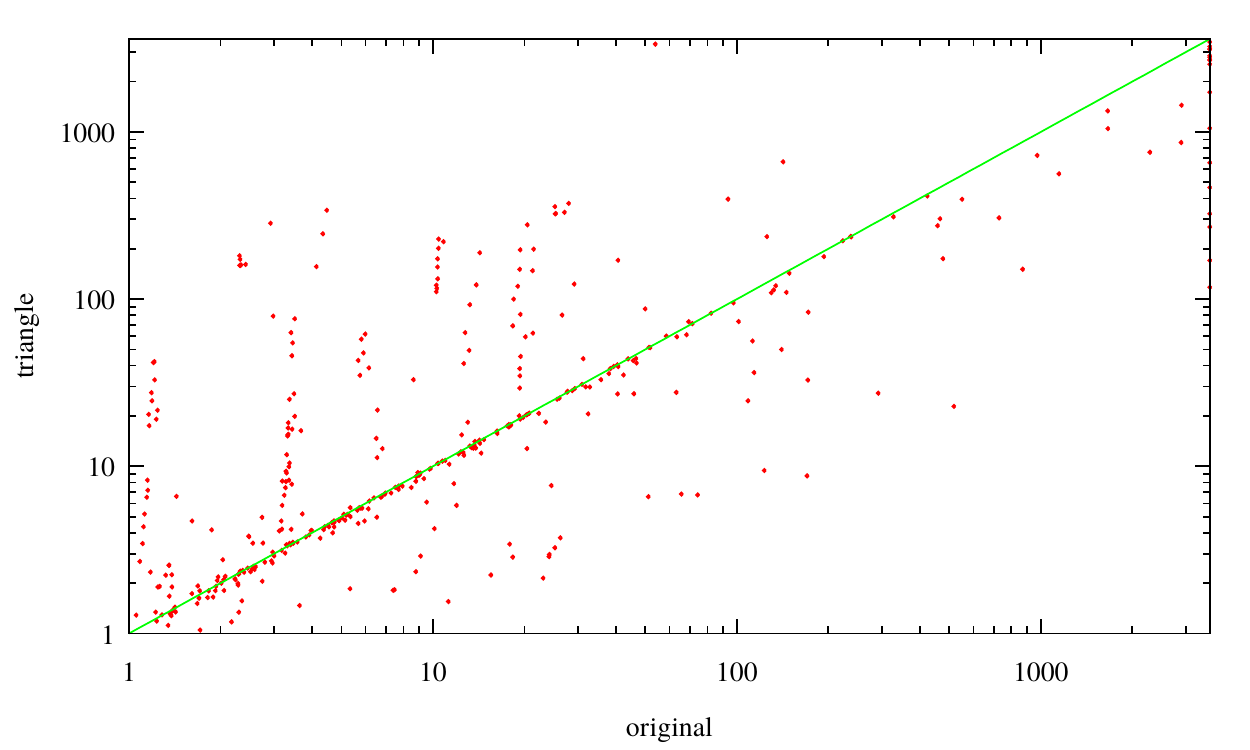}}}
\caption{Comparison of the runtime of MAC+RST+NG on the original instances and on the same instances after eliminating 
variables thanks to the triangle property.}
\label{original-triangle_solving_cumul_runtime}
\end{figure}

This subsection is devoted to the impact of elimination rules on the solving efficiency. 
So, we consider the 1,337 instances for which at least one of the considered rules allows to eliminate some 
variables. For each of them, we apply MAC+RST+NG on the original instance and on the instance 
after (possibly) eliminating variables and we compare the observed runtime.
In the latter case, the runtime includes both the solving runtime and the variable-elimination phase runtime.
Figure \ref{runtime_solving} gives the cumulative number of instances solved by MAC+RST+NG 
after eliminating some variables or by considering the original instances.
The ``step'' which appears after 30 minutes in the BT curve is due to the fact that a large number of instances
have just reached the time-out for the variable-elimination phase and are then solved fairly quickly.
To compare fairly the different algorithms we have to observe the curves after this 30-minute mark. 
It 
appears that MAC+RST+NG solves more instances when the instances are preprocessed 
with any of the considered elimination rules.
Moreover, the triangle property is again the most interesting elimination rule. 
Its use allows MAC+RST+NG to solve 1,008 instances while 
it only solves 996 and 998 instances when the instances are preprocessed respectively with 
the BT-degree property and $\exists$snake rule (1,000 instances for DE-snake rule).
Without any preprocessing, MAC+RST+NG performs worst by solving only 991 instances.
If we compare more finely the runtime of MAC+RST+NG applied on the original instances 
and on the instances after eliminating some variables thanks to the triangle property 
(see Figure \ref{original-triangle_solving_cumul_runtime}), we observe that, depending on the instance, 
eliminating variables may or may not improve solving efficiency.
However, we can remark that there exist several instances which MAC+RST+NG solves 
after the elimination of some variables but not without and above all that the converse is false.
So applying the triangle property for eliminating variables before solving makes sense.

Since the computational complexity of our variable-elimination rules is comparable with 
strong path consistency (SPC)~\cite{lecoutre}, it was natural to also test applying SPC.
However, applying SPC in preprocessing allowed us to solve only 869 instances 
compared to 991 instances using MAC+RST+NG alone (without any variable elimination) and hence
proved to be counter-productive. Applying SPC required the rewriting in extension of those relations
that need to be modified, whereas our variable-elimination rules allow us to keep these
relations in their original form.

\section{Variable-elimination rules and tractability}  \label{sec:tractability}

We investigate, in this section, the possibility of defining tractable classes based on our variable-elimination rules.
As is the case for BTP~\cite{Cooper10:btp}, the rules we have presented in this paper also define tractable classes
that can be detected in polynomial time by successive elimination of variables.

We also study the confluence of our variable-elimination rules which allows us to show the tractability
of maximising the number of eliminated variables.

\begin{definition} \label{def:order}
For a variable-elimination property $P$, we say that a binary CSP instance $I$ \emph{satisfies $P$ for the variable order $<$} if for
each variable $x_m$, except for the first variable according to the order $<$, $I$ satisfies the property $P$   
on $x_m$ in the sub-instance of $I$ restricted to the variables $x_i$ such that $x_i \leq x_m$.
\end{definition}

\begin{definition}
We say that a property $P$ of binary CSP instances is \emph{hereditary} if for any instance $I$
with more than one variable, $I$ satisfies $P$ implies that $I_{-m}$ satisfies $P$, where $I_{-m}$
is the instance obtained from $I$ after elimination of the variable $x_m$.
\end{definition}

\begin{theorem} \label{thm:tractablehereditary}
Let $P$ be an hereditary sol-var-elim property which can be tested in polynomial time.
The class of binary CSP instances $I$ satisfying the property $P$ (for a possibly unknown ordering of its variables)
can be detected and solved in polynomial time.
\end{theorem}

\begin{altproof}
Let $I$ be a binary CSP instance on $n$ variables.  
Suppose that $I$ satisfies the hereditary variable-elimination property $P$ for a variable ordering $<$.
Then $I$ satisfies property $P$ on the last variable of the ordering $<$. 
We can therefore find a variable $x_m$ on which $I$ satisfies the property $P$ by exhaustive search
over all $n$ variables. Note that there may be more than one variable which satisfies $P$.
In this case, we make an arbitrary choice which variable to eliminate: the
rest of the proof does require that $x_m$ be the last variable according to the order $<$.
Variable $x_m$ is then eliminated to produce the sub-instance $I_{-m}$
which has the same satisfiability as $I$. Since $P$ is hereditary, the instance $I_{-m}$ also satisfies $P$. 
By successive elimination
of variables we can reduce $I$ to an equivalent instance $I'$ on a single variable in polynomial time.
A single-variable instance being trivial to solve,
and since $P$ is a sol-var-elim property, we can construct a solution to $I$ in polynomial time. 
\end{altproof}

The following theorem is a direct consequence of Theorem~\ref{thm:tractablehereditary} and 
the fact that the listed properties are hereditary.

\begin{theorem} \label{tractable}
The class of binary CSP instances $I$ satisfying any of the following properties
(for a possibly unknown ordering of its variables) can be detected and solved in polynomial time:
\begin{enumerate}
\item the $\forall\exists$ broken $k$-dimensional polyhedron property (for any fixed $k \geq 2$),
\item the $\exists$snake property,
\item the DE-snake property,
\item $\forall\exists$BTP, 
\item the BT-degree property.
\end{enumerate}
\end{theorem}

At first sight, it might appear that we would not have an equivalent result for the triangle property, since
eliminating a variable $x_j$ might destroy the triangle property on another variable $x_i$ ($i \neq j$).
Recall that, in Definition~\ref{def:triangle}, a variable $x_i$ can be eliminated by the triangle property
only if there is a variable $x_j$ which justifies this elimination, so the obvious question is whether $x_i$
can still be eliminated after (its justifying variable) $x_j$ has been eliminated. It turns out that the answer
is yes, as we will now demonstrate.

\begin{definition}  \label{def:justifies}
In a binary CSP instance $I$, for distinct variable $x_i,x_j$, variable $x_j$ \emph{justifies} the elimination
by the triangle property 
of variable $x_i$ (which we denote $x_j \xrightarrow{\scriptstyle{I}} x_i$) if for all $v_j \in \mathcal{D}(x_j)$,
there exists $u_{ji}(v_j) \in \mathcal{D}(x_i)$ satisfying the following conditions:
\begin{description}
\item[ \ \ {\bf C1}($i$,$j$,$v_j$):] \ \ $(u_{ji}(v_j),v_j) \in R_{ij}$,
\item[ \ \ {\bf C2}($i$,$j$,$v_j$):] \ \ $\forall x_k \in X \setminus \{x_i,x_j\}$, $\forall v_k \in \mathcal{D}(x_k)$, \ 
$(v_j,v_k) \in R_{jk} \ \Rightarrow \ (u_{ji}(v_j),v_k) \in R_{ik}$.
\end{description}
\end{definition}

\begin{lemma}  \label{lem:almosttrans}
If $x_p \xrightarrow{\scriptstyle{I}} x_j$ and $x_j \xrightarrow{\scriptstyle{I}} x_i$, where $p \neq i$, 
then $x_p \xrightarrow{\scriptstyle{I_{-j}}} x_i$. In other words, if $x_p$ justifies the elimination of $x_j$
and $x_j$ justifies the elimination of $x_i$ by the triangle property, then $x_p$
justifies the elimination of $x_i$ in the instance $I_{-j}$ which is the result of the elimination of $x_j$ from $I$.
\end{lemma}

\begin{altproof}
Suppose that $x_p \xrightarrow{\scriptstyle{I}} x_j$ and $x_j \xrightarrow{\scriptstyle{I}} x_i$, where $p \neq i$.
For $v_p \in \mathcal{D}(x_p)$, define
\[ u_{pi}(v_p) \ := \ u_{ji}(u_{pj}(v_p)).
\]
It suffices to show that $u_{pi}(v_p)  \in \mathcal{D}(x_i)$ and that it satisfies the conditions of Definition~\ref{def:justifies}
in $I_{-j}$, namely:
\begin{description}
\item[ \ \ {\bf C1}($i$,$p$,$v_p$):] \ \ $(u_{pi}(v_p),v_p) \in R_{ip}$,
\item[ \ \ {\bf C2}($i$,$p$,$v_p$):] \ \ $\forall x_k \in X \setminus \{x_i,x_p,x_j\}$, $\forall v_k \in \mathcal{D}(x_k)$, \ 
$(v_p,v_k) \in R_{pk} \ \Rightarrow \ (u_{pi}(v_p),v_k) \in R_{ik}$.
\end{description}
Since $x_p \xrightarrow{\scriptstyle{I}} x_j$, we have $u_{pj}(v_p) \in \mathcal{D}(x_j)$ and then,
since $x_j \xrightarrow{\scriptstyle{I}} x_i$, we have  $u_{pi}(v_p) \in \mathcal{D}(x_i)$ (with $u_{pi}(v_p) = u_{ji}(u_{pj}(v_p))$).
Since $x_p \xrightarrow{\scriptstyle{I}} x_j$, we also have
from {\bf C1}($j$,$p$,$v_p$) that $(u_{pj}(v_p),v_p) \in R_{jp}$. 
Then, since $x_j \xrightarrow{\scriptstyle{I}} x_i$, we have 
{\bf C2}($i$,$j$,$u_{pj}(v_p)$) and, in particular for $k=p$ and $v_k=v_p \in \mathcal{D}(x_p)$:
\[ (u_{pj}(v_p),v_p) \in R_{jp} \ \Rightarrow \ (u_{ji}(u_{pj}(v_p)),v_p) \in R_{ip}
\]
Since $u_{pi}(v_p) = u_{ji}(u_{pj}(v_p))$, we can deduce that condition {\bf C1}($i$,$p$,$v_p$) holds.

Now consider any $x_k \in X \setminus \{x_i,x_p,x_j\}$ and any $v_k \in \mathcal{D}(x_k)$. 
Since $x_p \xrightarrow{\scriptstyle{I}} x_j$, we have from {\bf C2}($j$,$p$,$v_p$) that
\[ (v_p,v_k) \in R_{pk} \ \Rightarrow \ (u_{pj}(v_p),v_k) \in R_{jk}.
\]
Since $x_j \xrightarrow{\scriptstyle{I}} x_i$, we have from  {\bf C2}($i$,$j$,$u_{pj}(v_p)$):
\[ (u_{pj}(v_p),v_k) \in R_{jk} \ \Rightarrow \ (u_{ji}(u_{pj}(v_p)),v_k) \in R_{ik}. 
\]
Hence, we have 
\[ (v_p,v_k) \in R_{pk} \ \Rightarrow \ (u_{ji}(u_{pj}(v_p)),v_k) \in R_{ik}. 
\]
Since $u_{pi}(v_p) = u_{ji}(u_{pj}(v_p))$, it follows that condition {\bf C2}($i$,$p$,$v_p$) holds,
which completes the proof.
\end{altproof}

To complete our study of the definition of a tractable class based on the triangle property, we need to consider
the one case not covered by Lemma~\ref{lem:almosttrans}, namely $p=i$. For this, we require the notions
of isomorphic instances and neighbourhood substitutability of values. Indeed, when
$x_i \xrightarrow{\scriptstyle{I}} x_j$ and $x_j \xrightarrow{\scriptstyle{I}} x_i$ we have to choose which
of $x_i$ or $x_j$ we eliminate by the triangle property. We will show that modulo isomorphism (and provided
we have applied neighbourhood substitution), the resulting instances are identical and hence we can make an arbitrary
choice between $x_i$ and $x_j$.

\begin{definition}
Two binary CSP instances $I = \tuple{X^I,\mathcal{D}^I,R^I}$,  $J = \tuple{X^J,\mathcal{D}^J,R^J}$
are \emph{isomorphic} if there exist bijections $f : X^I \rightarrow X^J$ and 
$g_i : \mathcal{D}^I(x_i) \rightarrow \mathcal{D}^J(f(x_i))$ (for all $x_i \in X^I$) such that
for all pairs of distinct variables $x_i,x_j \in X^I$, for all $v_i \in \mathcal{D}^I(x_i)$ and for all $v_j \in \mathcal{D}^I(x_j)$,
\[ (v_i,v_j) \in R^I_{ij} \ \Leftrightarrow \ (g_i(v_i),g_j(v_j)) \in R^J_{f(i)f(j)}
\]
\end{definition}

A neighbourhood-substitutable value can be eliminated from its domain without changing the satisfiability
of the instance~\cite{DBLP:conf/aaai/Freuder91,DBLP:conf/gcai/FreuderW17}.

\begin{definition}
In a binary CSP instance  $I = \tuple{X,\mathcal{D},R}$,  $v_i \in \mathcal{D}(x_i)$ is
\emph{neighbourhood substitutable} by $v'_i \in \mathcal{D}(x_i) \setminus \{v_i\}$ if for all $x_j \in X \setminus \{x_i\}$,
for all $v_j \in \mathcal{D}(x_j)$,
\[ (v_i,v_j) \in R_{ij} \ \Rightarrow \ (v'_i,v_j) \in R_{ij}
\]
Two values $v_i$ and $v'_i$ are \emph{interchangeable}
if $v_i$ is neighbourhood substitutable by $v'_i$ and $v'_i$ is neighbourhood substitutable by $v_i$.
\end{definition}

Given a binary CSP instance $I$, it is known that the result of eliminating neighbourhood-substitutable values
until convergence (i.e. no more eliminations are possible) is unique up to isomorphism
~\cite{DBLP:journals/ai/Cooper97}. We assume that there is a program $\mathit{NS}$ which performs value eliminations
by neighbourhood substitutability until convergence, and we denote by $\mathit{NS}(I)$ the result of applying $\mathit{NS}$ to $I$.

\begin{lemma} \label{lem:isom}
If  $x_i \xrightarrow{\scriptstyle{I}} x_j$ and $x_j \xrightarrow{\scriptstyle{I}} x_i$ then $\mathit{NS}(I_{-i})$
and $\mathit{NS}(I_{-j})$ are isomorphic.
\end{lemma}

\begin{altproof}
It follows from the definition of $x_i \xrightarrow{\scriptstyle{I}} x_j$
that for all $v_i \in \mathcal{D}(x_i)$, we have $u_{ij}(v_i) \in \mathcal{D}(x_j)$
such that for all $x_k \in X \setminus \{x_i,x_j\}$, $\forall v_k \in \mathcal{D}(x_k)$,
\begin{equation} \label{eq:1}
(v_i,v_k) \in R_{ik} \ \Rightarrow \ (u_{ij}(v_i),v_k) \in R_{jk}
\end{equation}
Since $x_j \xrightarrow{\scriptstyle{I}} x_i$ we then have $u_{ji}(u_{ij}(v_i)) \in \mathcal{D}(x_i)$ such that
for all $x_k \in X \setminus \{x_i,x_j\}$, $\forall v_k \in \mathcal{D}(x_k)$,
\begin{equation} \label{eq:2}
(u_{ij}(v_i),v_k) \in R_{jk} \ \Rightarrow \ (u_{ji}(u_{ij}(v_i)),v_k) \in R_{ik}
\end{equation}
For each $x_i \in X$, define the function $F_i : \mathcal{D}(x_i)  \rightarrow \mathcal{D}(x_i)$ by
$F_i(v_i) = u_{ji}(u_{ij}(v_i))$ and consider the sequence
\[ v_i, F(v_i), F(F(v_i)), \ldots, F^r(v_i), \ldots
\]
Since $\mathcal{D}(x_i)$ is finite, this sequence must cycle at some point. Let $r \in \mathbb{N}$ be
the first value for which $F^r(v_i) = F^s(v_i)$ for some $s > r$. Thus, the above sequence has a cycle
of length $s-r$ starting at $F^r(v_i)$. 

From Equations~\ref{eq:1} and  \ref{eq:2}, we know that for all $x_k \in X \setminus \{x_i,x_j\}$,
$\forall v_k \in \mathcal{D}(x_k)$,
\[ (v_i,v_k) \in R_{ik} \ \Rightarrow \ (F(v_i),v_k) \in R_{ik}
\]
By a simple inductive argument, we can deduce that for any $t \in \mathbb{N}$,
\begin{equation} \label{eq:3}
(F^t(v_i),v_k) \in R_{ik} \ \Rightarrow \ (F^r(v_i),v_k) \in R_{ik}
\end{equation}
and hence each $F^t(v_i) \neq F^r(v_k)$ is neighbourhood substitutable by $F^r(v_i)$ in $I_{-j}$.
Thus, modulo isomorphism, we can assume that all values in the sequence
$v_i, F(v_i), F^2(v_i), \ldots$ have been eliminated from $\mathcal{D}(x_i)$ in $\mathit{NS}(I_{-j})$ by neighbourhood
substitution except for $F^r(v_i)$~\cite{DBLP:journals/ai/Cooper97}. By a similar argument,
modulo isomorphism, we can assume that in $\mathit{NS}(I_{-i})$ all values in the sequence
$u_{ij}(v_i), u_{ij}(F(v_i)), u_{ij}(F^2(v_i)), \ldots$ have been eliminated by neighbourhood
substitutability from $\mathcal{D}(x_j)$ except for $u_{ij}(F^r(v_i))$.
Furthermore, combining Equations~\ref{eq:1}, \ref{eq:2} and \ref{eq:3},  
we can deduce that for all $x_k \in X \setminus \{x_i,x_j\}$, $\forall v_k \in \mathcal{D}(x_k)$,
\[ (F^r(v_i),v_k) \in R_{ik} \ \Leftrightarrow \ (u_{ij}(F^r(v_i)),v_k) \in R_{jk}
\]
Thus, for each value in $\mathcal{D}(x_i)$ in $\mathit{NS}(I_{-j})$, there is a corresponding value
in $\mathcal{D}(x_j)$ in $\mathit{NS}(I_{-i})$ which has the same compatibilities with all values for
all other variables (and vice versa). Furthermore, no two values in $\mathcal{D}(x_i)$ 
(respectively, $\mathcal{D}(x_j)$) can have the same compatibilities with all values for all other variables, 
otherwise they would be interchangeable in $\mathit{NS}(I_{-j})$ (respectively, $\mathit{NS}(I_{-i})$) 
which would contradict the definition of neighbourhood substitution.
It follows that $\mathit{NS}(I_{-i})$ and $\mathit{NS}(I_{-j})$ are isomorphic.
\end{altproof}

We require one final lemma.

\begin{lemma} \label{lem:ns}
If $x_i \xrightarrow{\scriptstyle{I}} x_j$ then $x_i \xrightarrow{\scriptstyle{NS(I)}} x_j$.
\end{lemma}

\begin{altproof}
Suppose that $x_i \xrightarrow{\scriptstyle{I}} x_j$. We can see from Definition~\ref{def:triangle} that no 
eliminations of values from $\mathcal{D}(x_i)$ or $\mathcal{D}(x_k)$ (for $x_k \in X \setminus \{x_i,x_j\}$)
can possibly invalidate the elimination of $x_j$ by the triangle property. Suppose that $v_j \in \mathcal{D}(x_j)$
is eliminated in $NS(I)$ since it is neighbourhood substitutable by $v'_j \in \mathcal{D}(x_i)$. If $v_j = u_{ij}(v_i)$
for some $v_i \in \mathcal{D}(x_i)$, it suffices to set the value of $u_{ij}(v_i)$ to $v'_j$ instead of $v_j$.
It is easy to see that neighbourhood substitutability guarantees that conditions {\bf C1}($j$,$i$,$v_i$)
and {\bf C2}($j$,$i$,$v_i$) in Definition~\ref{def:justifies} hold with this new value of $u_{ij}(v_i)$.
\end{altproof}

We can now prove that the triangle property defines a tractable class which is detectable in polynomial time.

\begin{theorem} \label{thm:tractabletriangle}
The class of binary CSP instances $I$ satisfying the triangle property (for a possibly unknown ordering of its variables)
can be detected and solved in polynomial time.
\end{theorem}

\begin{altproof}
Suppose that there exists a variable order $<$ for which $I$ satisfies the triangle property. Let 
$x_m$ be the last variable according to this (unknown) order. We can find in $O(end^3)$, using the 
algorithm in Appendix~\ref{sec:apptriangle} the set $T$ of variables which could be eliminated from $I$
by (a first pass of) the triangle property. We know that $T \neq \emptyset$ since $x_m \in T$.
We do not know which variable in $T$ is the last variable according to the order $<$, so
we eliminate some arbitrary variable $x_i \in T$ from $I$. We then perform neighbourhood-substitution
eliminations until convergence to obtain $I' = NS(I_{-i})$. By Lemma~\ref{lem:almosttrans}, $x_m$
can be eliminated by the triangle property from $I'$ except possibly in the case that 
$x_i \xrightarrow{\scriptstyle{I}} x_m$ and $x_m \xrightarrow{\scriptstyle{I}} x_i$. But, in this latter case,
by Lemma~\ref{lem:isom}, $I'$ is isomorphic to $NS(I_{-m})$ and so it is as if we had eliminated $x_m$
instead of $x_i$. Lemma~\ref{lem:ns} tells us that that eliminating values by
neighbourhood substitutability does not destroy the fact that an instance satisfies the triangle property.
We can deduce that the instance $I'$ satisfies the triangle property and hence, by an easy inductive argument,
that we will reduce the instance to a single-variable instance by successive eliminations of $n-1$ variables.
The theorem follows from Theorem~\ref{prop:var-elim}.
\end{altproof}

When not all variables can be eliminated, we are interested in maximising the number of eliminated variables.
As pointed out in the proof of Theorem~\ref{thm:tractablehereditary}, 
the elimination of a variable by a hereditary rule cannot be invalidated by the elimination of another variable.
The following theorem is an immediate consequence of this and the fact that the listed properties are hereditary.

\begin{theorem}
Maximising the number of variables that can be eliminated by any of the following rules
can be achieved in polynomial time:
 the $\forall\exists$ broken $k$-dimensional polyhedron property (for any fixed $k \geq 2$),
 the $\exists$snake property,
 the DE-snake property,
 $\forall\exists$BTP and 
 the BT-degree property.
\end{theorem}

We saw in the proof of Theorem~\ref{thm:tractabletriangle} that the elimination of a variable
$x_m$ by the triangle property can only be invalidated by the elimination of another variable $x_i$
by the triangle property if $NS(I_{-i})$ is isomorphic to $NS(I_{-m})$. It follows that the triangle
property is confluent modulo isomorphism, provided neighbourhood substitution is applied after
every variable elimination. We thus have the following theorem.

\begin{theorem} Maximising the number of variable eliminations
by combining the triangle property and neighbourhood substitution can be achieved in polynomial time.
\end{theorem}

\section{Discussion and conclusion}

In this paper we have given novel satisfiability-conserving variable-elimination rules for binary CSPs,
two of which (namely DE-snake and BT-degree) strengthen previously-published rules.
In each case, if the instance is satisfiable, then a solution to the original instance can
be recovered in low-order polynomial time from a solution to the reduced instance. 
We have given optimised algorithms for applying each rule until convergence.
The DE-snake rule can be applied until convergence in $O(ed^3)$ time, whereas the corresponding time complexity 
for the triangle rule and the BT-degree rule is $O(end^3)$. 
However, it should be pointed out that the DE-snake rule inherits the disadvantage of the $\exists$snake
rule that the number of solutions may actually increase after elimination of a variable~\cite{ve}:
for example, it allows us to eliminate the central variable in the two-colouring of a star graph which
increases the number of solutions from $2$ to $2^{n-1}$.

Extensive experimental trials have confirmed that because of relatively high
time complexity of each of the variable-elimination rules, they may only be tested 
exhaustively during preprocessing. 
Applying them in preprocessing allowed us to solve more benchmark instances than without,
with the triangle rule allowing us to eliminate more variables and hence solve more instances than the other rules.
From a practical viewpoint, it would be interesting to understand how to better target the
instances or the variables for which the proposed variable elimination rules are likely to be profitable.
As a first step in this direction we have seen that most variables eliminated by our rules have small
domain size and/or small degree. Future work is required to determine whether versions of our rules
targetting only certain variables may be a practical possibility during search.

We have, in particular, generalised the notion of broken triangle to broken polyhedron,
which may be of independent theoretical interest.
The broken polyhedra property may lead to other possible theoretical advances 
(such as value-merging~\cite{OnBT-AIJ}, value-elimination~\cite{ve},
and generalisations to the general-arity CSP~\cite{OnBT-AIJ,DBLP:journals/constraints/Mouelhi17} 
or the Quantified CSP~\cite{DBLP:conf/aaai/GaoYZ11}), as was the case
with the broken-triangle property~\cite{Cooper10:btp}.

We have also shown that each of the variable-elimination rules allows us to define a novel hybrid 
tractable class by successive elimination of almost all variables. For each rule, 
this elimination order can be found in polynomial time,
which we found surprising in the case of the triangle property.

\vspace{1cm}

\acks{This work was funded by the Agence Nationale de la 
Recherche project ANR-16-C40-0028. 
}

\vspace{1cm}


\appendix

\section{Algorithm for variable elimination by the $\exists$snake property}  \label{sec:appsnake}

In this and the following appendices, we assume that we have implemented a set data structure in such a way that 
we can perform the following operations in $O(1)$ time: set membership, addition/deletion of an
element and testing whether the set is empty. This can be achieved using a boolean table together with 
a counter of the number of elements in the set, since in each case the set is a subset of a fixed set,
such as $X$ the variables of the instance.

Below we give an algorithm for eliminating variables by the  $\exists$snake property until convergence.
It uses the following data structures :
\begin{itemize}
\item $S_{\mathit{ELIM}}$ is the set of variables to be eliminated.
\item For $v_j,v'_j \! \in \! \mathcal{D}(x_j)$, vars$^{+}_{-}(j,v_j,v'_j) = \{ k \mid \exists v_k \! \in \! \mathcal{D}(x_k)
\text{ with } 
(v_j,v_k) \! \in \! R_{jk} \land (v'_j,v_k) \! \notin \! R_{jk} \}$.
\item For distinct $i,j \in \{1,\ldots,n\}$ such that $x_i$ constrains $x_j$ and $v_i \in \mathcal{D}(x_i)$,
countPairs($i,v_i,j$) is the number of pairs of values $v_j,v'_j \in \mathcal{D}(x_j)$ such that for some 
$k \neq i,j$ and some $v_k \in \mathcal{D}(x_k)$, the snake pattern (as shown in Figure~\ref{fig:VEsnake})
occurs on $\tuple{x_i,v_i}$, $\tuple{x_j,v_j}$, $\tuple{x_j,v'_j}$, $\tuple{x_k,v_k}$. We calculate 
countPairs($i,v_i,j$) by noting that it is the number of pairs of values $v_j,v'_j \in \mathcal{D}(x_j)$ such that 
 $(v'_j,v_i) \in R_{ji} \land (v_j,v_i) \notin R_{ji} \land$ vars$^{+}_{-}(j,v_j,v'_j) \setminus \{i\} \neq \emptyset$.
\item For $v_i \in \mathcal{D}(x_i)$, badVars($i,v_i$) is the set of $j \neq i$ such that countPairs($i,v_i,j$) $\neq 0$.
If  badVars($i,v_i$)$=\emptyset$, then variable $x_i$ is added to $S_{\mathit{ELIM}}$ .
\end{itemize}
The algorithm first initialises the above data structures, then performs eliminations from $X$,
the set of variables (which is initially $\{x_1,\ldots,x_n\}$).
When performing an elimination, the data structures are
updated which may lead to more variable eliminations.  Eliminations propagate until
convergence (i.e. until no more eliminations are possible). Updating the data structures, when a
variable $x_k$ is eliminated, means deleting $k$ from each vars$^{+}_{-}(j,v_j,v'_j)$ and each badVars($i,v_i$).
When deleting $k$ from vars$^{+}_{-}(j,v_j,v'_j)$, the value of countPairs($i,v_i,j$) needs to be
decremented only in the case that $((v'_j,v_i) \in R_{ji} \land (v_j,v_i) \notin R_{ji})$
and vars$^{+}_{-}(j,v_j,v'_j) \setminus \{i\}$ becomes empty for the first time (i.e. vars$^{+}_{-}(j,v_j,v'_j)$ becomes $\{i\}$
or it becomes empty and the $k$ being deleted from it is not $i$).

\begin{tabbing}
\hspace{5mm} \= \hspace{5mm} \= \hspace{5mm} \= \hspace{5mm} \= \hspace{5mm} \= \hspace{5mm} \= 
\hspace{5mm} \= \hspace{5mm} \=  \hspace{5mm} \= \hspace{5mm} \= \hspace{-60.5mm} 
*** Initialisation *** \\
$S_{\mathit{ELIM}}$ := $\emptyset$ ; \\
for $x_j \in X$ : \\
\> for $v_j \in \mathcal{D}(x_j)$ : \\
\> \> for $v'_j \in \mathcal{D}(x_j) \setminus \{v_j\}$ : \\
\> \> \> for $x_k \in X \setminus \{x_j\}$ such that $x_k$ is constrained by $x_j$ : \\
\> \> \> \> if $\exists v_k \in \mathcal{D}(x_k)$ such that $(v_j,v_k) \in R_{jk} \land (v'_j,v_k) \notin R_{jk}$ \\
\> \> \> \> then add $k$ to vars$^{+}_{-}(j,v_j,v'_j)$ ; \\ \\
for $x_i \in X$ : \\
\> for $v_i \in \mathcal{D}(x_i)$ : $..............................................................................................(1)$ \\
\> \> badVars($i,v_i$) := $\emptyset$ ; \\
\> \> for $x_j \in X \setminus \{x_i\}$ such that $x_j$ is constrained by $x_i$ : \\
\> \> \> countPairs($i,v_i,j$) := 0 ; \\
\> \> \> for $v_j \in \mathcal{D}(x_j)$ : \\
\> \> \> \> for $v'_j \in \mathcal{D}(x_j) \setminus \{v_j\}$ : \\
\> \> \> \> \> if $(v'_j,v_i) \in R_{ji} \land (v_j,v_i) \notin R_{ji} \land$ vars$^{+}_{-}(j,v_j,v'_j) \setminus \{i\} \neq \emptyset$ \\
\> \> \> \> \> then countPairs($i,v_i,j$) := countPairs($i,v_i,j$) $+1$ ; \\
\> \> \> \> \> \> badVars($i,v_i$) := badVars($i,v_i$) $\cup$ $\{j\}$ ; \\
\> \> if badVars($i,v_i$) $= \emptyset$ then add $x_i$ to $S_{\mathit{ELIM}}$ ; exit loop (1) ; \\
\\
*** Elimination and propagation *** \\
while $S_{\mathit{ELIM}} \neq \emptyset$ : \\
\> delete some $x_k$ from $S_{\mathit{ELIM}}$ ; \ $X$ : = $X \setminus \{x_k\}$ ; \\
\> for $x_j \in X$ such that $x_j$ is constrained by $x_k$ : \\
\> \> for $v_j \in \mathcal{D}(x_j)$ : \\
\> \> \> for $v'_j \in \mathcal{D}(x_j) \setminus \{v_j\}$ : \\
\> \> \> \> delete $k$ from vars$^{+}_{-}(j,v_j,v'_j)$ ; \\
\> \> \> \> if vars$^{+}_{-}(j,v_j,v'_j)$  becomes a singleton $\{i\}$ after deletion of $k$  $............(2)$ \\
\> \> \> \> then for $v_i \in \mathcal{D}(x_i)$ : \\
\> \> \> \> \> \> if $((v'_j,v_i) \in R_{ji} \land (v_j,v_i) \notin R_{ji})$ \\
\> \> \> \> \> \> then countPairs($i,v_i,j$) := countPairs($i,v_i,j$) $-1$ ; \\
\> \> \> \> \> \> \> if countPairs($i,v_i,j$) $=0$ \\
\> \> \> \> \> \> \> then badVars($i,v_i$) := badVars($i,v_i$) $\setminus \{j\}$ ; \\
\> \> \> \> \> \> \> \> if badVars($i,v_i$) $= \emptyset$ then add $x_i$ to $S_{\mathit{ELIM}}$ ; \\
\> \> \> \> if vars$^{+}_{-}(j,v_j,v'_j)$  becomes $\emptyset$ after deletion of $k$  $...............................(3)$ \\
\> \> \> \> then for $x_i \in X \setminus \{x_j,x_k\}$ such that $x_i$ is constrained by $x_j$ : \\
\> \> \> \> \> \> for $v_i \in \mathcal{D}(x_i)$ : \\
\> \> \> \> \> \> \> if $((v'_j,v_i) \in R_{ji} \land (v_j,v_i) \notin R_{ji})$ \\
\> \> \> \> \> \> \> then countPairs($i,v_i,j$) := countPairs($i,v_i,j$) $-1$ ; \\
\> \> \> \> \> \> \> \> if countPairs($i,v_i,j$) $=0$ \\
\> \> \> \> \> \> \> \> then badVars($i,v_i$) := badVars($i,v_i$) $\setminus \{j\}$ ; \\
\> \> \> \> \> \> \> \> \> if badVars($i,v_i$) $= \emptyset$ then add $x_i$ to $S_{\mathit{ELIM}}$ ; \\
\> for $i \in X$ such that $x_i$ is constrained by $x_k$ : \\
\> \> for $v_i \in \mathcal{D}(x_i)$ : $........................................................................................(4)$ \\
\> \> \> badVars($i,v_i$) := badVars($i,v_i$) $\setminus$ $\{k\}$ ; \\
\> \> \> if badVars($i,v_i$) $= \emptyset$ then add $x_i$ to $S_{\mathit{ELIM}}$ ; exit loop (4) ;
\end{tabbing}

This algorithm requires $O(ed^3)$ time and $O(ed^2)$ space. To see the  $O(ed^3)$ time bound,
observe that each of the tests (2) and (3) can only be True once for each triple $(j,v_j,v'_j)$.
The data structure vars$^{+}_{-}(j,v_j,v'_j)$ requires $O(ed^2)$ space.

\section{Algorithm for variable elimination by the DE-snake property}  \label{sec:appDEsnake}

Below we give an algorithm for eliminating variables by the DE-snake property until convergence.
It uses the following data structures :
\begin{itemize}
\item $S_{\mathit{ELIM}}$ is the set of variables to be eliminated.
\item For $v_j,v'_j \! \in \! \mathcal{D}(x_j)$, vars$^{+}_{-}(j,v_j,v'_j) = \{ k \mid \exists v_k \! \in \! \mathcal{D}(x_k)
\text{ with } 
(v_j,v_k) \! \in \! R_{jk} \land (v'_j,v_k) \! \notin \! R_{jk} \}$.
\item For $v_i \in \mathcal{D}(x_i)$, badAssts($i,v_i$) is the set of assignments $\tuple{j,v_j}$ such that
$(v_i,v_j) \notin R_{ij}$ and $\nexists v'_j \in \mathcal{D}(x_j)$ such that
vars$^{+}_{-}(j,v_j,v'_j) \subseteq \{i\}$ and $(v_i,v'_j) \in R_{ij}$. 
If  badAssts($i,v_i$)$=\emptyset$, then variable $x_i$ is added to $S_{\mathit{ELIM}}$ since it can be elminated by the 
DE-snake rule.
\end{itemize}
The algorithm first initialises the above data structures, then performs eliminations from $X$,
the set of variables (which is initially $\{x_1,\ldots,x_n\}$).
When performing an elimination, the data structures are
updated which may lead to more variable eliminations.  Eliminations propagate until
convergence (i.e. until no more eliminations are possible). Updating the data structures, when a
variable $x_k$ is eliminated, means deleting $k$ from each vars$^{+}_{-}(j,v_j,v'_j)$ and 
deleting each assignment to $x_k$ from each badAssts($i,v_i$).
When deleting $k$ from vars$^{+}_{-}(j,v_j,v'_j)$, badAssts($i,v_i$) needs to be updated
for each variable $x_i$ for which vars$^{+}_{-}(j,v_j,v'_j) \setminus \{i\}$ becomes empty for the first time
(i.e. vars$^{+}_{-}(j,v_j,v'_j)$ becomes $\{i\}$ or it becomes empty and the $k$ being deleted from it is not $i$).

\begin{tabbing}
\hspace{5mm} \= \hspace{5mm} \= \hspace{5mm} \= \hspace{5mm} \= \hspace{5mm} \= \hspace{5mm} \= 
\hspace{5mm} \= \hspace{5mm} \=  \hspace{5mm} \= \hspace{5mm} \= \hspace{-60.5mm} 
*** Initialisation *** \\
$S_{\mathit{ELIM}}$ := $\emptyset$ ; \\
for $x_j \in X$ : \\
\> for $v_j \in \mathcal{D}(x_j)$ : \\
\> \> for $v'_j \in \mathcal{D}(x_j) \setminus \{v_j\}$ : \\
\> \> \> for $x_k \in X \setminus \{x_j\}$ such that $x_k$ is constrained by $x_j$ : \\
\> \> \> \> if $\exists v_k \in \mathcal{D}(x_k)$ such that $(v_j,v_k) \in R_{jk} \land (v'_j,v_k) \notin R_{jk}$ \\
\> \> \> \> then add $k$ to vars$^{+}_{-}(j,v_j,v'_j)$ ; \\
for $x_i \in X$ : \\
\> for $v_i \in \mathcal{D}(x_i)$ : \\
\> \> badAssts($i,v_i$)  := $\emptyset$ \\
\> \> for $x_j$ such that $x_j$ is constrained by $x_i$ : \\
\> \> \> for $v_j \in \mathcal{D}(x_j)$ : \\
\> \> \> \> if $(v_i,v_j) \notin R_{ij}$ \\
\> \> \> \> then for $v'_j \in \mathcal{D}(x_j) \setminus \{v_j\}$ : \\
\> \> \> \> \> \> if $(v_i,v'_j) \in R_{ij}$ and vars$^{+}_{-}(j,v_j,v'_j) \setminus \{i\} = \emptyset$ \\
\> \> \> \> \> \> then add  $\{\tuple{j,v_j}\}$ to badAssts($i,v_i$) ; \\
\> \> if badAssts($i,v_i$) $= \emptyset$ then add $x_i$ to $S_{\mathit{ELIM}}$ ;  \\
\\
\\ \\ \\ \\ \\ \\
*** Elimination and propagation *** \\
while $S_{\mathit{ELIM}} \neq \emptyset$ : \\
\> delete some $x_k$ from $S_{\mathit{ELIM}}$ ; \ $X$ : = $X \setminus \{x_k\}$ ; \\
\> for $x_j \in X$ such that $x_j$ is constrained by $x_k$ : \\
\> \> for $v_j \in \mathcal{D}(x_j)$ : \\
\> \> \> for $v'_j \in \mathcal{D}(x_j) \setminus \{v_j\}$ : \\
\> \> \> \> delete $k$ from vars$^{+}_{-}(j,v_j,v'_j)$ ; \\
\> \> \> \> if vars$^{+}_{-}(j,v_j,v'_j)$  becomes a singleton $\{i\}$ after deletion of $k$ : $...........(1)$ \\
\> \> \> \> then for $v_i \in \mathcal{D}(x_i)$ : \\
\> \> \> \> \> \> if $((v'_j,v_i) \in R_{ji} \land (v_j,v_i) \notin R_{ji})$  \\
\> \> \> \> \> \> then badAssts($i,v_i$) := badAssts($i,v_i$) $\setminus \{\tuple{j,v_j}\}$ ; \\
\> \> \> \> \> \> \> if badAssts($i,v_i$) becomes $\emptyset$ then add $x_i$ to $S_{\mathit{ELIM}}$ ; \\
\> \> \> \> if vars$^{+}_{-}(j,v_j,v'_j)$  becomes $\emptyset$ after deletion of $k$ : $..............................(2)$ \\
\> \> \> \> then for $x_i \in X \setminus \{x_j,x_k\}$ such that $x_i$ is constrained by $x_j$ : \\
\> \> \> \> \> \> for $v_i \in \mathcal{D}(x_i)$ : \\
\> \> \> \> \> \> \> if $((v'_j,v_i) \in R_{ji} \land (v_j,v_i) \notin R_{ji})$ \\
\> \> \> \> \> \> \> then badAssts($i,v_i$) := badAssts($i,v_i$) $\setminus \{\tuple{j,v_j}\}$ ; \\
\> \> \> \> \> \> \> \> if badAssts($i,v_i$) becomes $\emptyset$ then add $x_i$ to $S_{\mathit{ELIM}}$ ; \\
\> for $i \in X$ such that $x_i$ is constrained by $x_k$ : \\
\> \> for $v_i \in \mathcal{D}(x_i)$ : \\
\> \> \> for $v_k \in \mathcal{D}(x_k)$ : \\
\> \> \> \> badAssts($i,v_i$) := badVars($i,v_i$) $\setminus$ $\{\tuple{k,v_k}\}$ ; \\
\> \> \> \> if badVars($i,v_i$) becomes $\emptyset$ then add $x_i$ to $S_{\mathit{ELIM}}$ ; 
\end{tabbing}

This algorithm requires $O(ed^3)$ time and $O(ed^2)$ space. 
To see the $O(ed^3)$ time bound, observe that each of the tests (1) and (2) can only be True once for each triple $(j,v_j,v'_j)$.
The data structures vars$^{+}_{-}(j,v_j,v'_j)$ and badAssts($i,v_i$) both require $O(ed^2)$ space.

\section{Algorithm for variable elimination by the triangle property}  \label{sec:apptriangle}

Below we give an algorithm for eliminating variables by the triangle property until convergence.
It uses the following data structures :
\begin{itemize}
\item $S_{\mathit{ELIM}}$ is the set of variables to be eliminated.
\item For $v_j \in \mathcal{D}(x_j)$, $v_i \in \mathcal{D}(x_i)$ such that $(v_j,v_i) \in R_{ji}$,
badVars($j,v_j,i,v_i$) is the set of $k \neq i,j$ such that there exists $v_k \in \mathcal{D}(x_k)$ with
$(v_j,v_k) \in R_{jk}$ and $(v_i,v_k) \notin R_{ik}$.
\item supported($j,v_j,i$) = True if there exists $v_i \in \mathcal{D}(x_i)$ such that
badVars($j,v_j,i,v_i$) = $\emptyset$.
\item count($j,i$) is the number of values $v_j \in \mathcal{D}(x_j)$ such that supported($j,v_j,i$) = False.
If count($j,i$) = 0, for some $j \neq i$ such that $x_j \in X \setminus S_{\mathit{ELIM}}$ 
then we can eliminate $x_i$ by the triangle property.
\end{itemize}
The algorithm first initialises the above data structures, then performs eliminations.
When performing an elimination, these data structures are
updated which may lead to further variable eliminations. The only delicate point in the 
algorithm is that once a variable is due to be eliminated it cannot be used to justify the 
elimination of another variable; hence 
the test $x_j \notin S_{\mathit{ELIM}}$ 
in lines (1) and (3).

\begin{tabbing}
\hspace{5mm} \= \hspace{5mm} \= \hspace{5mm} \= \hspace{5mm} \= \hspace{5mm} \= \hspace{5mm} \= 
\hspace{5mm} \= \hspace{5mm} \=  \hspace{5mm} \= \hspace{5mm} \= \hspace{-60.5mm} 
*** Initialisation *** \\
$S_{\mathit{ELIM}}$ := $\emptyset$ ; \\
for $x_i \in X$ : \\
\> for $x_j \in X \setminus \{x_i\}$ such that $x_j \notin S_{\mathit{ELIM}}$ : $......................................................(1)$ \\
\> \> count($j,i$) := 0 ; \\
\> \> for $v_j \in \mathcal{D}(x_j)$ \\
\> \> \> supported($j,v_j,i$) := False ; \\
\> \> \> for $v_i \in \mathcal{D}(x_i)$ such that $(v_j,v_i) \in R_{ji}$ : $..............................................(2)$ \\
\> \> \> \> badVars($j,v_j,i,v_i$) := $\emptyset$ ; \\
\> \> \> \> for $x_k \in X \setminus \{x_i,x_j\}$ such that $x_k$ is constrained by $x_i$ : \\
\> \> \> \> \> if $\exists v_k \in \mathcal{D}(x_k)$ such that $(v_j,v_k) \in R_{jk}$ and $(v_i,v_k) \notin R_{ik}$ \\
\> \> \> \> \> then add $k$ to badVars($j,v_j,i,v_i$) ; \\
\> \> \> \> if badVars($j,v_j,i,v_i$) = $\emptyset$ \\
\> \> \> \> then supported($j,v_j,i$) := True ; \ exit loop (2) ; \\
\> \> \> if supported($j,v_j,i$) = False then count($j,i$) := count($j,i$) $+1$ ; \\
\> \> if count($j,i$) = 0 \\
\> \> then add $x_i$ to $S_{\mathit{ELIM}}$ ; \ exit loop (1) ; \\
\\
*** Elimination and propagation *** \\
while $S_{\mathit{ELIM}} \neq \emptyset$ : \\
\> delete some $x_k$ from $S_{\mathit{ELIM}}$ ; \ $X$ : = $X \setminus \{x_k\}$ ;\\
\> for $x_i \in X$ such that $x_i$ is constrained by $x_k$ : \\
\> \> for $x_j \in X \setminus \{x_i\}$  such that $x_j \notin S_{\mathit{ELIM}}$ : $...............................................(3)$ \\
\> \> \> for $v_j \in \mathcal{D}(x_j)$ such that supported($j,v_j,i$) = False : \\
\> \> \> \> for $v_i \in \mathcal{D}(x_i)$ : $..........................................................................(4)$ \\
\> \> \> \> \> if $k \in$ badVars($j,v_j,i,v_i$) \\
\> \> \> \> \> then delete $k$ from badVars($j,v_j,i,v_i$) \\
\> \> \> \> \> \> if badVars($j,v_j,i,v_i$) = $\emptyset$ \\
\> \> \> \> \> \> then supported($j,v_j,i$) := True ; \\
\> \> \> \> \> \> \> count($j,i$) := count($j,i$) $-1$ ; \\
\> \> \> \> \> \> \> if count($j,i$) = 0 \\
\> \> \> \> \> \> \> then add $x_i$ to $S_{\mathit{ELIM}}$ ; \ exit loop (3) ; \\
\> \> \> \> \> \> \> exit loop (4) ;
\end{tabbing}

This algorithm requires $O(end^3)$ time and $O(end^2)$ space. Curiously, the
propagation phase requires less time than initialisation phase, since it requires only $O(end^2)$ time.
This is because we no longer need to look at individual values in the propagation phase.

\section{Algorithm for variable elimination by the BT-degree property}  \label{sec:appBTdegree}

Below we give an algorithm for eliminating variables by the  BT-degree property until convergence.
It uses the following data structures :
\begin{itemize}
\item $S_{\mathit{ELIM}}$ is the set of variables to be eliminated.
\item $N^{+}_{-}(i,v_i,j,v_j,m) = \  \big| \{ v_m \in \mathcal{D}(x_m) 
\mid (v_i,v_m) \in R_{im} \land (v_j,v_m) \notin R_{jm} \} \big|$,
is the number of values $v_m \in \mathcal{D}(x_m)$ which are linked by a positive edge to $\tuple{x_i,v_i}$ and by a 
negative edge to $\tuple{x_j,v_j}$.
\item For a pair of variables $x_i,x_m$ linked by a constraint, for $v_i \in \mathcal{D}(x_i)$ and 
$v_m \in \mathcal{D}(x_m)$, BTvars($i,v_i,m,v_m$) is the set of $j \neq i,m$ such that there is a broken
triangle $(v_m,v_i,v_j,v'_m)$ or $(v_m,v_j,v_i,v'_m)$ for some $v_j \in \mathcal{D}(x_j)$ and $v'_m \in \mathcal{D}(x_m)$.
\item BTdegree($i,v_i,m,v_m$) is the cardinality of the set BTvars($i,v_i,m,v_m$).
\item 3safe($i,v_i,j,v_j,m$) is True if $(v_i,v_j) \in R_{ij}$ is 3-safe on $x_m$.
\item $M^{+}_{-}(i,v_i,j,v_j,m)$ is similar to  $N^{+}_{-}(i,v_i,j,v_j,m)$, 
except that the negative edge $v_jv_m$ must also have BT degree greater than 1.
\item badBases($m$) is the set of $(i,v_i,j,v_j)$ such that $i,j,m$ are distinct, $v_i \in \mathcal{D}(x_i)$,
$v_j \in \mathcal{D}(x_j)$ and $\nexists v'_m \in \mathcal{D}(x_m)$ satisfying the conditions in the definition of the 
BT-degree property (namely, $(v_i,v'_m) \in R_{im}$, $(v_j,v'_m) \in R_{jm}$ and either $(v_i,v_j)$ is 3-safe on $x_m$
or $(v_i,v'_m)$ has BT degree zero or $(v_j,v'_m)$ has BT degree zero). Thus, if  badBases($m$) is the empty set, then
$x_m$ can be eliminated by the  BT-degree property.
\end{itemize}
The algorithm first initialises the above data structures, then performs eliminations.
Again, the set of variables $X$ is initially $\{x_1,\ldots,x_n\}$. The data structure 3safe is calculated using the fact that
 $(v_i,v_j) \in R_{ij}$ is 3-safe on $x_m$ if and only if $M^{+}_{-}(i,v_i,j,v_j,m) = 0$ or 
$M^{+}_{-}(j,v_j,i,v_i,m) = 0$.
When performing an elimination, all data structures are updated which may provoke further variable eliminations. 

\begin{tabbing}
\hspace{4mm} \= \hspace{4mm} \= \hspace{4mm} \= \hspace{4mm} \= \hspace{4mm} \= \hspace{4mm} \= 
\hspace{4mm} \= \hspace{4mm} \=  \hspace{4mm} \= \hspace{4mm} \= \hspace{-50.5mm} 
*** Initialisation *** \\
$S_{\mathit{ELIM}}$ := $\emptyset$ ; \\
for $x_m \in X$ : \\
\> for $x_i \in X \setminus \{x_m\}$ such that $x_i$ is constrained by $x_m$: \\
\> \> for $x_j \in X \setminus \{x_i,x_m\}$ such that $x_j$ is constrained by $x_m$ : \\
\> \> \> for $v_i \in \mathcal{D}(x_i)$ : \\
\> \> \> \> for $v_j \in \mathcal{D}(x_j)$ such that $(v_i,v_j) \in R_{ij}$ : \\
\> \> \> \> \> $N^{+}_{-}(i,v_i,j,v_j,m)$ := $\big| \{ v_m \in \mathcal{D}(x_m) \mid (v_i,v_m) \in R_{im} 
\land (v_j,v_m) \notin R_{jm} \} \big|$ ; \\
for $x_m \in X$ : \\
\> for $x_i \in X \setminus \{x_m\}$ such that $x_i$ is constrained by $x_m$: \\
\> \> for $v_i \in \mathcal{D}(x_i)$ : \\
\> \> \> for $v_m \in \mathcal{D}(x_m)$ : \\
\> \> \> \> BTvars($i,v_i,m,v_m$) := $\{ j \neq i,m \mid \exists v_j \in \mathcal{D}(x_j)$ such that $(v_i,v_j) \in R_{ij}$ and \\
\> \> \> \> \> \> ($(v_i,v_m) \notin R_{im} \land (v_j,v_m) \in R_{jm} \land N^{+}_{-}(i,v_i,j,v_j,m) \neq 0$) \\
\> \> \> \> \> \> or  ($(v_i,v_m) \in R_{im} \land (v_j,v_m) \notin R_{jm} \land N^{+}_{-}(j,v_j,i,v_i,m) \neq 0$) $\}$ ; \\
\> \> \> \> BTdegree($i,v_i,m,v_m$) := $\mid$BTvars($i,v_i,m,v_m$)$\mid$ ; \\
for $x_m \in X$ : \\
\> badBases($m$) := $\emptyset$ ; \\
\> for $x_i \in X \setminus \{x_m\}$ such that $x_i$ is constrained by $x_m$: \\
\> \> for $x_j \in X \setminus \{x_i,x_m\}$ such that $x_j$ is constrained by $x_m$ : \\
\> \> \> for $v_i \in \mathcal{D}(x_i)$ : \\
\> \> \> \> for $v_j \in \mathcal{D}(x_j)$ such that $(v_i,v_j) \in R_{ij}$ : $.........................................................(1)$ \\
\> \> \> \> \> $M^{+}_{-}(i,v_i,j,v_j,m)$ :=  $\big| \{ v_m \in \mathcal{D}(x_m) \mid (v_i,v_m) \in R_{im} \land (v_j,v_m) \notin R_{jm}$ \\
\> \> \> \> \> \hspace{3cm} $\land$ BTdegree($j,v_j,m,v_m$) $>1 \} \big|$ ; \\
\> \> \> \> \> 3safe($i,v_i,j,v_j,m$) := $M^{+}_{-}(i,v_i,j,v_j,m) = 0$ or $M^{+}_{-}(j,v_j,i,v_i,m) = 0$ ; \\
\> \> \> \> \> if $\nexists v'_m \in \mathcal{D}(x_m)$ such that $(v_i,v'_m) \in R_{im} \land (v_j,v'_m)\in R_{jm}$ $\land$ \\
\> \> \> \>  \> \>  (3safe($i,v_i,j,v_j,m$) $\lor$ BTdegree($i,v_i,m,v'_m$) = 0 $\lor$ BTdegree($j,v_j,m,v'_m$) = 0) \\
\> \> \> \> \> then add ($i,v_i,j,v_j$) to badBases($m$) ; \ exit loop (1) ; \\
\> if badBases($m$) = $\emptyset$  then add $x_m$ to $S_{\mathit{ELIM}}$ ; \\
\\
\\
*** Elimination and propagation *** \\
while $S_{\mathit{ELIM}} \neq \emptyset$ : \\
\> delete some $x_j$ from $S_{\mathit{ELIM}}$ ; \ $X$ : = $X \setminus \{x_j\}$ ; \\
\> for $x_m \in X$ such that $x_m$ is constrained by $x_j$ : \\
\> \> for $x_i \in X \setminus \{x_m\}$ such that $x_i$ is constrained by $x_m$ : \\
\> \> \> for $v_i \in \mathcal{D}(x_i)$ : \\
\> \> \> \> for $v_m \in \mathcal{D}(x_m)$ : \\
\> \> \> \> \> if $j \in$ BTvars($i,v_i,m,v_m$) \\
\> \> \> \> \> then delete $j$ from BTvars($i,v_i,m,v_m$) ; \\
\> \> \> \> \> \> BTdegree($i,v_i,m,v_m$) := BTdegree($i,v_i,m,v_m$) $-1$ ; \\
\> \> \> \> \> \> if BTdegree($i,v_i,m,v_m$) $=1$  $..............................................................(2)$ \\
\> \> \> \> \> \> then for $x_k \in X \setminus \{x_i,x_m\}$ such that $x_k$ is constrained by $x_m$ : \\
\> \> \> \> \> \> \> \> for $v_k \in \mathcal{D}(x_k)$ such that $(v_i,v_k) \in R_{ik}$ : \\
\> \> \> \> \> \> \> \> \> if $(v_k,v_m) \in R_{km} \land (v_i,v_m) \notin R_{im}$ \\
\> \> \> \> \> \> \> \> \> then $M^{+}_{-}(k,v_k,i,v_i,m)$ := $M^{+}_{-}(k,v_k,i,v_i,m) - 1$ ; \\
\> \> \> \> \> \> \> \> \> if $(M^{+}_{-}(i,\!v_i,\!k,\!v_k,\!m) = 0 \lor M^{+}_{-}(k,\!v_k,\!i,\!v_i,\!m) = 0)$ becomes True \\
\> \> \> \> \> \> \> \> \> then 3safe($i,v_i,k,v_k,m$) := True ; \\ 
\> \> \> \> \> \> \> \> \> \> for $v'_m \in \mathcal{D}(x_m)$ such that $(v_i,\!v'_m) \in R_{im} \land (v_k,\!v'_m) \in R_{km}$ : \\
\> \> \> \> \> \> \> \> \> \> \hspace{5mm}\= badBases($m$) := badBases($m$) $\setminus \{ (i,v_i,k,v_k) \}$ ; \\
\> \> \> \> \> \> \> \> \> \> \> if badBases($m$) = $\emptyset$ then add $x_m$ to $S_{\mathit{ELIM}}$ ; \\
\> \> \> \> \> \> if BTdegree($i,v_i,m,v_m$) $= 0$  $..............................................................(3)$ \\
\> \> \> \> \> \> then for $x_k \in X \setminus \{x_i,x_m\}$ such that $x_k$ is constrained by $x_m$ ; \\
\> \> \> \> \> \> \> \> for $v_k \in \mathcal{D}(x_k)$ such that $(v_i,v_k) \in R_{ik}$ : \\
\> \> \> \> \> \> \> \> \> if $(v_i,v_m) \in R_{im}$ and $(v_k,v_m) \in R_{km}$ \\
\> \> \> \> \> \> \> \> \> then badBases($m$) := badBases($m$) $\setminus \{ (i,v_i,k,v_k) \}$ ; \\
\> \> \> \> \> \> \> \> \> \> if badBases($m$) = $\emptyset$ then  add $x_m$ to $S_{\mathit{ELIM}}$ ; 
\end{tabbing}

This algorithm requires $O(end^3)$ time and $O(end^2 \log d)$ space. To prove the $O(end^3)$ time bound we have to
use the fact that each of the tests (2) and (3) can only become True once for each quadruple $(i,v_i,m,v_m)$.
The data structure badBases requires $O(end^2)$ space and the
data structures $N^{+}_{-}$ 
and $M^{+}_{-}$ 
require $O(end^2 \log d)$ space,
the $\log d$ factor being due to the fact that they store integers up to $d$.

\section{Algorithm for variable elimination by $\forall\exists$BTP}  \label{sec:appBTP}

Below we give an algorithm for eliminating variables by $\forall\exists$BTP until convergence.
It uses the following data structures :
\begin{itemize}
\item $S_{\mathit{ELIM}}$ is the set of variables to be eliminated.
\item badVars($i$) is the set of $j \neq i$ such that there is some value $v_j \in \mathcal{D}(x_j)$
such that for all values $v_i \in \mathcal{D}(x_i)$ there is a broken triangle $(v_i,v_j,v_k,v'_i)$ on $x_i$ for some $v'_i,v_k,k$ such
that $v'_i \in \mathcal{D}(x_i)$ and $v_k \in \mathcal{D}(x_k)$.
\item countBadVals($j$,$i$) is the number of values $v_j \in \mathcal{D}(x_j)$ such that
for all values $v_i \in \mathcal{D}(x_i)$ there is a broken triangle $(v_i,v_j,v_k,v'_i)$ on $x_i$ for some $v'_i,v_k,k$ such
that $v'_i \in \mathcal{D}(x_i)$ and $v_k \in \mathcal{D}(x_k)$.
\item  support($j$,$v_j$,$i$) is the set of all values $v_i \in \mathcal{D}(x_i)$ such that 
there is no broken triangle
 $(v_i,v_j,v_k, v'_i)$ on $x_i$ for any $v'_i,v_k,k$ with $v'_i \in \mathcal{D}(x_i)$ and $v_k \in \mathcal{D}(x_k)$.
\item $L_{BT}(j,v_j,i,v_i)$ is the set of $k$ such that there is a broken triangle $(v_i,v_j,v_k,v'_i)$ on $x_i$ for some $v'_i,v_k,k$ 
such that $v'_i \in \mathcal{D}(x_i)$ and $v_k \in \mathcal{D}(x_k)$.
\end{itemize}
The algorithm first initialises the above data structures, then performs eliminations.
When performing an elimination, these data structures are updated which may provoke further variable eliminations. 

\begin{tabbing}
\hspace{5mm} \= \hspace{5mm} \= \hspace{5mm} \= \hspace{5mm} \= \hspace{5mm} \= \hspace{5mm} \= 
\hspace{5mm} \= \hspace{5mm} \=  \hspace{5mm} \= \hspace{5mm} \= \hspace{-60.5mm} 
*** Initialisation *** \\
$S_{\mathit{ELIM}}$ := $\emptyset$ ; \\
for $x_i \in X$ : \\
\> badVars($i$) := $\emptyset$ ;  \\
\> for $x_j \in X \setminus \{x_i\}$ such that $x_j$ is constrained by $x_i$ : \\
\> \> countBadVals($j$,$i$) := 0 ; \\
\> \> for $v_j \in \mathcal{D}(x_j)$ : \\
\> \> \> support($j$,$v_j$,$i$) := $\emptyset$ ; \ $L$ := $\emptyset$ ; \\
\> \> \> for $x_k \in X \setminus \{x_i,x_j\}$ such that $x_k$ is constrained by $x_i$ : \\
\> \> \> \> for $v_k \in \mathcal{D}(x_k)$ such that $(v_j,v_k) \in R_{jk}$ : \\
\> \> \> \> \> if $\exists v'_i \in \mathcal{D}(x_i)$ such that $(v_j,v'_i) \notin R_{ji}$ and $(v'_i,v_k) \in R_{ik}$ \\
\> \> \> \> \> then add $\tuple{k,v_k}$ to $L$ ; \\
\> \> \> for $v_i \in \mathcal{D}(x_i)$ such that $(v_j,v_i) \in R_{ji}$ : \\
\> \> \> \> $L_{BT}(j,v_j,i,v_i)$ := $\emptyset$ ; \\
\> \> \> \> for $\tuple{k,v_k} \in L$ : \\
\> \> \> \> \> if $(v_i,v_k) \notin R_{ik}$ then $L_{BT}(j,v_j,i,v_i)$ := $L_{BT}(j,v_j,i,v_i) \cup \{k\}$ ; \\
\> \> \> \> if $L_{BT}(j,v_j,i,v_i)$ = $\emptyset$ then add $v_i$ to support($j$,$v_j$,$i$) ; \\
\> \> \> if support($j$,$v_j$,$i$) = $\emptyset$ \\
\> \> \> then countBadVals($j$,$i$) := countBadVals($j$,$i$) $+1$ ; \ badVars($i$) := badVars($i$) $\cup$ $\{j\}$ ; \\
\> if badVars($i$) = $\emptyset$ then add $x_i$ to $S_{\mathit{ELIM}}$ ; \\ 
\\
\\ \\
*** Elimination and propagation *** \\
while $S_{\mathit{ELIM}} \neq \emptyset$ : \\
\> delete some variable $x_k$ from $S_{\mathit{ELIM}}$ ;  \ $X$ := $X \setminus \{x_k\}$ ; \\
\> for $x_i \in X$ such that $x_i$ is constrained by $x_k$ : \\
\> \> if $k$ $\in$ badVars($i$) then delete $k$ from badVars($i$) ; \\
\> \> for $x_j \in X \setminus \{x_i,x_k\}$ such that $x_j$ is constrained by $x_i$ : \\
\> \> \> for $v_j \in \mathcal{D}(x_j)$ : \\
\> \> \> \> for $v_i \in \mathcal{D}(x_i)$ such that $(v_j,v_i) \in R_{ji}$ : \\
\> \> \> \> \> if $k \in L_{BT}(j,v_j,i,v_i)$ \\
\> \> \> \> \> then delete $k$ from $L_{BT}(j,v_j,i,v_i)$ ; \\
\> \> \> \> \> \> if $L_{BT}(j,v_j,i,v_i) = \emptyset$ \\
\> \> \> \> \> \> then add $v_i$ to support($j$,$v_j$,$i$) ; \\
\> \> \> \> \> \> \> if $|$support($j$,$v_j$,$i$)$|$ = 1 \\
\> \> \> \> \> \> \> then countBadVals($j$,$i$) := countBadVals($j$,$i$) $ - 1$ ; \\
\> \> \> \> \> \> \> \> if countBadVals($j$,$i$) $= 0$ then delete $j$ from badVars($i$) ; \\
\> \> if badVars($i$) $= \emptyset$ then add $x_i$ to $S_{\mathit{ELIM}}$ ;
\end{tabbing}

This algorithm requires $O(end^3)$ time and $O(end^2)$ space. The data structure $L_{BT}$ requires  $O(end^2)$ space.

\vskip 0.2in

\bibliography{biblio}

\begin{thebibliography}{}

\bibitem[\protect\BCAY{Bassiliades, Bikakis, Vrakas, Vlahavas,\ \BBA\
  Vouros}{Bassiliades et~al.}{2016}]{DBLP:conf/setn/2016}
Bassiliades, N., Bikakis, A., Vrakas, D., Vlahavas, I.~P., \BBA\ Vouros,
  G.~A.\BEDS. \BBOP2016\BBCP.
\newblock {\Bem Proceedings of the 9th Hellenic Conference on Artificial
  Intelligence, {SETN} 2016, Thessaloniki, Greece, 2016}. {ACM}.

\bibitem[\protect\BCAY{Beigel\ \BBA\ Eppstein}{Beigel\ \BBA\
  Eppstein}{1995}]{DBLP:conf/focs/BeigelE95}
Beigel, R.\BBACOMMA\  \BBA\ Eppstein, D. \BBOP1995\BBCP.
\newblock \BBOQ {3-Coloring in Time O(1.3446\({}^{\mbox{n}}\)): {A} No-{MIS}
  Algorithm}\BBCQ\
\newblock In {\Bem 36th Annual Symposium on Foundations of Computer Science,
  Milwaukee, Wisconsin, USA}, \BPGS\ 444--452. {IEEE} Computer Society.

\bibitem[\protect\BCAY{Bessi{\`{e}}re, R{\'{e}}gin, Yap,\ \BBA\
  Zhang}{Bessi{\`{e}}re et~al.}{2005}]{ac}
Bessi{\`{e}}re, C., R{\'{e}}gin, J., Yap, R. H.~C., \BBA\ Zhang, Y.
  \BBOP2005\BBCP.
\newblock \BBOQ An optimal coarse-grained arc consistency algorithm\BBCQ\
\newblock {\Bem Artif. Intell.}, {\Bem 165\/}(2), 165--185.

\bibitem[\protect\BCAY{Boussemart, Hemery, Lecoutre,\ \BBA\ Sais}{Boussemart
  et~al.}{2004}]{DBLP:conf/ecai/BoussemartHLS04}
Boussemart, F., Hemery, F., Lecoutre, C., \BBA\ Sais, L. \BBOP2004\BBCP.
\newblock \BBOQ {Boosting Systematic Search by Weighting Constraints}\BBCQ\
\newblock In {\Bem Proceedings of the 16th Eureopean Conference on Artificial
  Intelligence, ECAI 2004}, \BPGS\ 146--150.

\bibitem[\protect\BCAY{Carbonnel, Cohen, Cooper,\ \BBA\ Zivny}{Carbonnel
  et~al.}{2018}]{stacs18}
Carbonnel, C., Cohen, D.~A., Cooper, M.~C., \BBA\ Zivny, S. \BBOP2018\BBCP.
\newblock \BBOQ {On Singleton Arc Consistency for {CSP}s Defined by Monotone
  Patterns}\BBCQ\
\newblock In Niedermeier, R.\BBACOMMA\  \BBA\ Vall{\'{e}}e, B.\BEDS, {\Bem 35th
  Symposium on Theoretical Aspects of Computer Science, {STACS} 2018, Caen,
  France}, \lowercase{\BVOL}~96 of {\Bem LIPIcs}, \BPGS\ 19:1--19:15. Schloss
  Dagstuhl - Leibniz-Zentrum fuer Informatik.

\bibitem[\protect\BCAY{Cohen, Cooper, Escamocher,\ \BBA\ Zivny}{Cohen
  et~al.}{2015}]{ve}
Cohen, D.~A., Cooper, M.~C., Escamocher, G., \BBA\ Zivny, S. \BBOP2015\BBCP.
\newblock \BBOQ Variable and value elimination in binary constraint
  satisfaction via forbidden patterns\BBCQ\
\newblock {\Bem J. Comput. Syst. Sci.}, {\Bem 81\/}(7), 1127--1143.

\bibitem[\protect\BCAY{Cooper}{Cooper}{1997}]{DBLP:journals/ai/Cooper97}
Cooper, M.~C. \BBOP1997\BBCP.
\newblock \BBOQ Fundamental properties of neighbourhood substitution in
  constraint satisfaction problems\BBCQ\
\newblock {\Bem Artif. Intell.}, {\Bem 90\/}(1-2), 1--24.

\bibitem[\protect\BCAY{Cooper}{Cooper}{2014}]{beyond}
Cooper, M.~C. \BBOP2014\BBCP.
\newblock \BBOQ {Beyond Consistency and Substitutability}\BBCQ\
\newblock In O'Sullivan, B.\BED, {\Bem Principles and Practice of Constraint
  Programming - 20th International Conference, {CP} 2014, Lyon, France},
  \lowercase{\BVOL}\ 8656 of {\Bem Lecture Notes in Computer Science}, \BPGS\
  256--271. Springer.

\bibitem[\protect\BCAY{Cooper, Duchein, {El Mouelhi}, Escamocher, Terrioux,\
  \BBA\ Zanuttini}{Cooper et~al.}{2016a}]{OnBT-AIJ}
Cooper, M.~C., Duchein, A., {El Mouelhi}, A., Escamocher, G., Terrioux, C.,
  \BBA\ Zanuttini, B. \BBOP2016a\BBCP.
\newblock \BBOQ {Broken triangles: From value merging to a tractable class of
  general-arity constraint satisfaction problems}\BBCQ\
\newblock {\Bem Artif. Intell.}, {\Bem 234}, 196--218.

\bibitem[\protect\BCAY{Cooper, {El Mouelhi},\ \BBA\ Terrioux}{Cooper
  et~al.}{2016b}]{wBTP}
Cooper, M.~C., {El Mouelhi}, A., \BBA\ Terrioux, C. \BBOP2016b\BBCP.
\newblock \BBOQ {Extending Broken Triangles and Enhanced Value-Merging}\BBCQ\
\newblock In Rueher, M.\BED, {\Bem Principles and Practice of Constraint
  Programming - 22nd International Conference, {CP} 2016, Toulouse, France},
  \lowercase{\BVOL}\ 9892 of {\Bem Lecture Notes in Computer Science}, \BPGS\
  173--188. Springer.

\bibitem[\protect\BCAY{Cooper, Jeavons,\ \BBA\ Salamon}{Cooper
  et~al.}{2010}]{Cooper10:btp}
Cooper, M.~C., Jeavons, P.~G., \BBA\ Salamon, A.~Z. \BBOP2010\BBCP.
\newblock \BBOQ {Generalizing constraint satisfaction on trees: Hybrid
  tractability and variable elimination}\BBCQ\
\newblock {\Bem Artif. Intell.}, {\Bem 174\/}(9-10), 570--584.

\bibitem[\protect\BCAY{Cooper, J{\'{e}}gou,\ \BBA\ Terrioux}{Cooper
  et~al.}{2015}]{kbtp}
Cooper, M.~C., J{\'{e}}gou, P., \BBA\ Terrioux, C. \BBOP2015\BBCP.
\newblock \BBOQ {A Microstructure-Based Family of Tractable Classes for
  {CSP}s}\BBCQ\
\newblock In Pesant, G.\BED, {\Bem Principles and Practice of Constraint
  Programming - 21st International Conference, {CP} 2015, Cork, Ireland},
  \lowercase{\BVOL}\ 9255 of {\Bem Lecture Notes in Computer Science}, \BPGS\
  74--88. Springer.

\bibitem[\protect\BCAY{Cooper\ \BBA\ Zivny}{Cooper\ \BBA\
  Zivny}{2017}]{DBLP:conf/dagstuhl/CooperZ17}
Cooper, M.~C.\BBACOMMA\  \BBA\ Zivny, S. \BBOP2017\BBCP.
\newblock \BBOQ {Hybrid Tractable Classes of Constraint Problems}\BBCQ\
\newblock In Krokhin, A.~A.\BBACOMMA\  \BBA\ Zivny, S.\BEDS, {\Bem The
  Constraint Satisfaction Problem: Complexity and Approximability},
  \lowercase{\BVOL}~7 of {\Bem Dagstuhl Follow-Ups}, \BPGS\ 113--135. Schloss
  Dagstuhl - Leibniz-Zentrum fuer Informatik.

\bibitem[\protect\BCAY{Dechter}{Dechter}{1999}]{DBLP:journals/ai/Dechter99}
Dechter, R. \BBOP1999\BBCP.
\newblock \BBOQ {Bucket Elimination: {A} Unifying Framework for
  Reasoning}\BBCQ\
\newblock {\Bem Artif. Intell.}, {\Bem 113\/}(1-2), 41--85.

\bibitem[\protect\BCAY{Dechter\ \BBA\ Pearl}{Dechter\ \BBA\
  Pearl}{1989}]{dechter89:tree}
Dechter, R.\BBACOMMA\  \BBA\ Pearl, J. \BBOP1989\BBCP.
\newblock \BBOQ {Tree Clustering for Constraint Networks}\BBCQ\
\newblock {\Bem Artif. Intell.}, {\Bem 38\/}(3), 353--366.

\bibitem[\protect\BCAY{{El Mouelhi}}{{El
  Mouelhi}}{2017}]{DBLP:journals/constraints/Mouelhi17}
{El Mouelhi}, A. \BBOP2017\BBCP.
\newblock \BBOQ Tractable classes for {CSP}s of arbitrary arity: {F}rom theory
  to practice\BBCQ\
\newblock {\Bem Constraints}, {\Bem 22\/}(1), 97--98.

\bibitem[\protect\BCAY{{El Mouelhi}}{{El
  Mouelhi}}{2018}]{DBLP:journals/constraints/Mouelhi18}
{El Mouelhi}, A. \BBOP2018\BBCP.
\newblock \BBOQ On a new extension of {BTP} for binary {CSP}s\BBCQ\
\newblock {\Bem Constraints}, {\Bem 23\/}(4), 355--382.

\bibitem[\protect\BCAY{Freuder}{Freuder}{1991}]{DBLP:conf/aaai/Freuder91}
Freuder, E.~C. \BBOP1991\BBCP.
\newblock \BBOQ Eliminating interchangeable values in constraint satisfaction
  problems\BBCQ\
\newblock In Dean, T.~L.\BBACOMMA\  \BBA\ McKeown, K.~R.\BEDS, {\Bem
  Proceedings of the 9th National Conference on Artificial Intelligence,
  Anaheim, CA, USA, 1991, Volume 1.}, \BPGS\ 227--233. {AAAI} Press / The {MIT}
  Press.

\bibitem[\protect\BCAY{Freuder\ \BBA\ Wallace}{Freuder\ \BBA\
  Wallace}{2017}]{DBLP:conf/gcai/FreuderW17}
Freuder, E.~C.\BBACOMMA\  \BBA\ Wallace, R.~J. \BBOP2017\BBCP.
\newblock \BBOQ Replaceability and the substitutability hierarchy for
  constraint satisfaction problems\BBCQ\
\newblock In Benzm{\"{u}}ller, C., Lisetti, C.~L., \BBA\ Theobald, M.\BEDS,
  {\Bem {GCAI} 2017, 3rd Global Conference on Artificial Intelligence, Miami,
  FL, USA}, \lowercase{\BVOL}~50 of {\Bem EPiC Series in Computing}, \BPGS\
  51--63. EasyChair.

\bibitem[\protect\BCAY{Gao, Yin,\ \BBA\ Zhou}{Gao
  et~al.}{2011}]{DBLP:conf/aaai/GaoYZ11}
Gao, J., Yin, M., \BBA\ Zhou, J. \BBOP2011\BBCP.
\newblock \BBOQ Hybrid tractable classes of binary quantified constraint
  satisfaction problems\BBCQ\
\newblock In Burgard, W.\BBACOMMA\  \BBA\ Roth, D.\BEDS, {\Bem Proceedings of
  the Twenty-Fifth {AAAI} Conference on Artificial Intelligence, {AAAI} 2011,
  San Francisco, USA}. {AAAI} Press.

\bibitem[\protect\BCAY{Jeavons, Cohen,\ \BBA\ Cooper}{Jeavons
  et~al.}{1998}]{DBLP:journals/ai/JeavonsCC98}
Jeavons, P., Cohen, D.~A., \BBA\ Cooper, M.~C. \BBOP1998\BBCP.
\newblock \BBOQ {Constraints, Consistency and Closure}\BBCQ\
\newblock {\Bem Artif. Intell.}, {\Bem 101\/}(1-2), 251--265.

\bibitem[\protect\BCAY{Koubarakis}{Koubarakis}{2006}]{DBLP:reference/fai/Koubarakis06}
Koubarakis, M. \BBOP2006\BBCP.
\newblock \BBOQ Temporal {CSPs}\BBCQ\
\newblock In Rossi, F., van Beek, P., \BBA\ Walsh, T.\BEDS, {\Bem Handbook of
  Constraint Programming}, \lowercase{\BVOL}~2 of {\Bem Foundations of
  Artificial Intelligence}, \BPGS\ 665--697. Elsevier.

\bibitem[\protect\BCAY{Kratsch, Philip,\ \BBA\ Ray}{Kratsch
  et~al.}{2016}]{DBLP:journals/talg/KratschPR16}
Kratsch, S., Philip, G., \BBA\ Ray, S. \BBOP2016\BBCP.
\newblock \BBOQ {Point Line Cover: The Easy Kernel is Essentially Tight}\BBCQ\
\newblock {\Bem {ACM} Trans. Algorithms}, {\Bem 12\/}(3), 40:1--40:16.

\bibitem[\protect\BCAY{Larrosa\ \BBA\ Dechter}{Larrosa\ \BBA\
  Dechter}{2003}]{DBLP:journals/constraints/LarrosaD03}
Larrosa, J.\BBACOMMA\  \BBA\ Dechter, R. \BBOP2003\BBCP.
\newblock \BBOQ {Boosting Search with Variable Elimination in Constraint
  Optimization and Constraint Satisfaction Problems}\BBCQ\
\newblock {\Bem Constraints}, {\Bem 8\/}(3), 303--326.

\bibitem[\protect\BCAY{Lecoutre}{Lecoutre}{2009}]{lecoutre}
Lecoutre, C. \BBOP2009\BBCP.
\newblock {\Bem {Constraint Networks Techniques and Algorithms}}.
\newblock ISTE/Wiley.

\bibitem[\protect\BCAY{Lecoutre, Sais, Tabary,\ \BBA\ Vidal}{Lecoutre
  et~al.}{2007}]{DBLP:journals/jsat/LecoutreSTV07}
Lecoutre, C., Sais, L., Tabary, S., \BBA\ Vidal, V. \BBOP2007\BBCP.
\newblock \BBOQ {Recording and Minimizing Nogoods from Restarts}\BBCQ\
\newblock {\Bem {JSAT}}, {\Bem 1\/}(3-4), 147--167.

\bibitem[\protect\BCAY{Naanaa}{Naanaa}{2013}]{DBLP:journals/jetai/Naanaa13}
Naanaa, W. \BBOP2013\BBCP.
\newblock \BBOQ Unifying and extending hybrid tractable classes of {CSPs}\BBCQ\
\newblock {\Bem J. Exp. Theor. Artif. Intell.}, {\Bem 25\/}(4), 407--424.

\bibitem[\protect\BCAY{Naanaa}{Naanaa}{2016}]{wadySETN}
Naanaa, W. \BBOP2016\BBCP.
\newblock \BBOQ Extending the {B}roken {T}riangle {P}roperty tractable class of
  binary {CSP}s\BBCQ.
\newblock In Bassiliades et~al. \cite{DBLP:conf/setn/2016}, \BPGS\ 3:1--3:6.

\bibitem[\protect\BCAY{Newman, Fr{\'{e}}chette,\ \BBA\ Leyton{-}Brown}{Newman
  et~al.}{2018}]{deepOpt}
Newman, N., Fr{\'{e}}chette, A., \BBA\ Leyton{-}Brown, K. \BBOP2018\BBCP.
\newblock \BBOQ Deep optimization for spectrum repacking\BBCQ\
\newblock {\Bem Commun. {ACM}}, {\Bem 61\/}(1), 97--104.

\bibitem[\protect\BCAY{Omrani\ \BBA\ Naanaa}{Omrani\ \BBA\
  Naanaa}{2016}]{DBLP:conf/setn/OmraniN16}
Omrani, M.~A.\BBACOMMA\  \BBA\ Naanaa, W. \BBOP2016\BBCP.
\newblock \BBOQ A constrained molecular graph generation with imposed and
  forbidden fragments\BBCQ.
\newblock In Bassiliades et~al. \cite{DBLP:conf/setn/2016}, \BPGS\ 4:1--4:5.

\bibitem[\protect\BCAY{Rossi, Petrie,\ \BBA\ Dhar}{Rossi
  et~al.}{1990}]{rossi90:equivalence}
Rossi, F., Petrie, C.~J., \BBA\ Dhar, V. \BBOP1990\BBCP.
\newblock \BBOQ {On the Equivalence of Constraint Satisfaction Problems}\BBCQ\
\newblock In {\Bem {ECAI}}, \BPGS\ 550--556.

\bibitem[\protect\BCAY{Schrijver}{Schrijver}{1999}]{Schrijver}
Schrijver, A. \BBOP1999\BBCP.
\newblock {\Bem Theory of Linear and Integer Programming}.
\newblock Wiley-Interscience Series in Discrete Mathematics and Optimization.
  Wiley.

\bibitem[\protect\BCAY{Subbarayan\ \BBA\ Pradhan}{Subbarayan\ \BBA\
  Pradhan}{2004}]{DBLP:conf/sat/SubbarayanP04}
Subbarayan, S.\BBACOMMA\  \BBA\ Pradhan, D.~K. \BBOP2004\BBCP.
\newblock \BBOQ {N}i{VER}: {N}on-increasing {V}ariable {E}limination
  {R}esolution for {P}reprocessing {SAT} instances\BBCQ\
\newblock In {\Bem {SAT} 2004 - The 7th International Conference on Theory and
  Applications of Satisfiability Testing, Vancouver, Canada, Online
  Proceedings}.

\bibitem[\protect\BCAY{Zhang\ \BBA\ Yap}{Zhang\ \BBA\
  Yap}{2011}]{DBLP:journals/tplp/ZhangY11}
Zhang, Y.\BBACOMMA\  \BBA\ Yap, R. H.~C. \BBOP2011\BBCP.
\newblock \BBOQ Solving functional constraints by variable substitution\BBCQ\
\newblock {\Bem {TPLP}}, {\Bem 11\/}(2-3), 297--322.

\end{thebibliography}
\bibliographystyle{theapa}

\end{document}